\documentclass[12pt,preprint]{aastex}

\usepackage{graphicx,amsmath,color,natbib,morefloats}

\begin{document}

\title{The Milky Way's Circular Velocity Curve to 60 kpc and an
  Estimate of the Dark Matter Halo Mass from the Kinematics of
  $\sim$2400 SDSS Blue Horizontal-Branch Stars}

\author{X.-X.  Xue\altaffilmark{1,2,3}, H.-W.  Rix\altaffilmark{2}, G.
  Zhao\altaffilmark{1}, P.  Re Fiorentin\altaffilmark{2,4}, T.
  Naab\altaffilmark{5}, M. Steinmetz\altaffilmark{7},\\ F. C. van den
  Bosch\altaffilmark{2}, T.  C.  Beers\altaffilmark{6},Y.  S.
  Lee\altaffilmark{6}, E.  F.  Bell\altaffilmark{2}, C.
  Rockosi\altaffilmark{8}, B.  Yanny\altaffilmark{9}, \\H.
  Newberg\altaffilmark{10}, R.  Wilhelm\altaffilmark{11}, X.
  Kang\altaffilmark{2}, M.  C.  Smith\altaffilmark{12}, D.  P.
  Schneider\altaffilmark{13}} \altaffiltext{1}{The National
  Astronomical Observatories, CAS, 20A Datun Road, Chaoyang District,
  100012, Beijing, China} \altaffiltext{2}{Max-Planck-Institute for
  Astronomy K\"{o}nigstuhl 17, D-69117, Heidelberg, Germany}
\altaffiltext{3}{Graduate University of the Chinese Academy of
  Sciences, 19A Yuquan Road, Shijingshan District, 100049, Beijing,
  China} \altaffiltext{4}{Department of Physics, University of
  Ljubljana, Jadranska 19, 1000 Ljubljana, Slovenia}
\altaffiltext{5}{Universit\"{a}t-Sternwarte M\"{u}nchen, Scheinerstr.
  1, D-81679, M\"{u}nchen, Germany} \altaffiltext{6}{Department of
  Physics and Astronomy, CSCE: Center for the Study of Cosmic
  Evolution, and JINA: Joint Institute for Nuclear Astrophysics,
  Michigan State University, E. Lansing, MI 48824, USA}
\altaffiltext{7}{Astrophysical Institute Potsdam, An der Sternwarte
  16, 14482 Potsdam, Germany.}  \altaffiltext{8}{Lick
  Observatory/University of California, Santa Cruz, CA 95060, USA}
\altaffiltext{9}{Fermi National Accelerator Laboratory, P.O.  Box 500
  Batavia, IL 60510-5011, USA} \altaffiltext{10}{Department of
  Physics, Applied Physics, and Astronomy, Rensselaer Polytechnic
  Institute, Troy, NY 12180, USA} \altaffiltext{11}{Department of
  Physics and Astronomy, Texas Tech University, Lubbock, TX 79409,
  USA} \altaffiltext{12}{Institute of Astronomy, University of
  Cambridge, Madingley Road, Cambridge. CB3 0HA, United Kingdom}
\altaffiltext{13}{Department of Astronomy and Astrophysics, 504 Davey
  Laboratory University Park, Pennsylvania 16802, USA}
%%%%%%%%%%%%%%%%%%%%%%%%%%%%%%%%%%%%%%%%%%%%%%%%%%%%%%%%%%%%%%%%%%%%%%%%%%%%%%%%%%%%%%%%%%%

\begin{abstract}

  We derive new constraints on the mass of the Milky Way's dark matter
  halo, based on a set of halo stars from SDSS as kinematic tracers.
  Our sample comprises $2401$ rigorously selected Blue
  Horizontal-Branch (BHB) halo stars at $\rm |z| \ge 4$ kpc, and with
  distances from the Galactic center up to $\sim 60$ kpc, with
  photometry and spectra drawn from SDSS DR-6. With distances accurate
  to $\sim 10\%$, this sample enables construction of the full
  line-of-sight velocity distribution at different Galactocentric
  radii. To interpret these distributions, we compare them to matched
  mock observations drawn from two different cosmological galaxy
  formation simulations designed to resemble the Milky Way, which we
  presume to have an appropriate orbital distribution of halo stars.
  Specifically, we select simulated halo stars in the same volume as
  the observations, and derive the distributions $\rm
  P(V_{los}/V_{cir})$ of their line-of-sight velocities for different
  radii, normalized by the simulation's local circular velocity. We
  then determine which value of $\rm V_{cir}(r)$ brings the observed
  distribution into agreement with the corresponding distributions
  from the simulations. These values are then adopted as observational
  estimates for $\rm V_{cir}(r)$, after a small Jeans Equation
  correction is made to account for slight data/simulation differences
  in the radial density distribution. This procedure results in an
  estimate of the Milky Way's circular velocity curve to $\sim
  60$~kpc, which is found to be slightly falling from the adopted
  value of $\rm 220~km~s^{-1}$ at the Sun's location, and implies
  M$(<60 \rm~ kpc) = 4.0\pm 0.7\times 10^{11}$M$_\odot$. The radial
  dependence of $\rm V_{cir}(r)$, derived in statistically independent
  bins, is found to be consistent with the expectations from an NFW
  dark matter halo with the established stellar mass components at its
  center. If we assume an NFW halo profile of characteristic
  concentration holds, we can use the observations to estimate the
  virial mass of the Milky Way's dark matter halo, M$_{\rm
    vir}=1.0^{+0.3}_{-0.2} \times 10^{12}$M$_\odot$, which is lower
  than many previous estimates. We have checked that the particulars
  of the cosmological simulations are unlikely to introduce
  systematics larger than the statistical uncertainties.  This
  estimate implies that nearly 40\% of the baryons within the virial
  radius of the Milky Way's dark matter halo reside in the stellar
  components of our Galaxy. A value for M$_{\rm vir}$ of only $\sim
  1\times10^{12}$M$_\odot$ also (re-)opens the question of whether all
  of the Milky Way's satellite galaxies are on bound orbits.

\end{abstract}


%%%%%%%%%%%%%%%%%%%%%%%%%%%%%%%%%%%%%%%%%%%%%%%%%%%%%%%%%%%%%%%%%%%%%%%%%%%%%%%%%%%%%%%%%%%%

\section{Introduction}

The visible parts of galaxies are, in the current paradigm for galaxy
formation, concentrations of baryons at the center of much larger dark
matter halos, which have assembled through hierarchical merging and
gas cooling. Understanding the properties of these dark matter host
halos, their virial masses, concentrations, and radial mass profiles,
{\it vis-a-vis} the luminous properties of the main galaxy at their
center, is crucial for modeling the dynamics of the central galaxy,
for connecting observations of galaxies to large-scale cosmological
dark matter simulations, and for understanding what fraction of
baryons in the halo ended up as stars in the central galaxy. In turn,
the extended stellar distributions of galaxies, in particular the
stellar halos of nearby galaxies, offer some of the best probes to
test generic predictions about the nature of dark matter mass profiles
(Navarro et al. 1996).

The Milky Way and its surrounding halo are of particular interest, as
our internal position permits the placement of unique constraints on
the Galaxy's stellar mass content, on its dark matter profile at large
radii, and on the 3-D shape of its dark matter halo. Yet, our location
within the Galaxy also complicates some measurements, such as the
extended rotation curve of gas in its disk. As a result, the dark mass
profile for the Milky Way between $\sim 10$ kpc and $\sim 100$ kpc and
the halo's virial mass have not been previously constrained to better
than a factor of 2 to 3. In practice, it has proven useful to quantify
the halo mass profile by either a circular velocity curve, $\rm
V_{cir}(r)$, or by the escape velocity curve, $\rm V_{esc}(r)$.

In previous work, the most common tools used to estimate the Milky Way
halo mass are the escape velocity and the velocity dispersion profile
of the tracer populations (i.e., halo stars, or the Milky Way's
satellite galaxies and globular clusters). The escape velocity
provides constraints on the gravitational potential at the relevant
position (Little \& Tremaine 1987; Zaritsky et al.  1989; Kulessa \&
Lynden-Bell 1992; Kochanek 1996). Recent work has shown that the total
mass of the halo is around $2 \times10^{12} ~ {\rm M}_\odot$.
Wilkinson \& Evans (1999) used the velocities of 27 satellite galaxies
and globular clusters to find a halo mass of $\sim 1.9^{+3.6}_{-1.7}
\times 10^{12}~{\rm M}_\odot$ by adopting a truncated, flat rotation
curve halo model.  Sakamoto, Chiba \& Beers (2003) used a sample
including 11 satellite galaxies, 137 globular clusters, and 413 solar
neighborhood field horizontal-branch stars, along with a flat rotation
curve model, to obtain a total halo mass of $2.5^{+0.5}_{-1.0} \times
10^{12}~{\rm M}_\odot$ or $1.8^{+0.4}_{-0.7} \times 10^{12}~{\rm
  M}_\odot$, depending on the inclusion (or not) of Leo I. Recently,
Smith et al. (2007) estimated a halo mass of $\sim
1.42^{+1.14}_{-0.54} \times 10^{12}~{\rm M}_\odot$, based on a sample
of high-velocity stars from the RAVE survey and two published
databases, and an adiabatically contracted NFW halo model. Battaglia
et al. (2005, 2006) used a derived velocity dispersion profile to
determine a total mass of $0.5 \sim 1.5 \times 10^{12}~{\rm M}_\odot$,
with some dependence on the model adopted for the halo profile (see
also Dehnen, McLaughlin \& Sachania 2006).

In order to improve the precision of the mass estimate of the halo,
the fundamental first step is to begin with high-quality data (tracers
with accurate distances and radial velocities), augmented with an
efficient method of analysis capable of extracting the maximum amount
of information. For example, Wilkinson \& Evans (1999) comment that a
data set of $\sim 200$ radial velocities of BHB stars reduces the
uncertainty in the mass estimate of the halo to $\sim 20$\%.  This
goal has proven elusive, even with data samples twice this size (e.g.
Sakamoto et al. 2003), due to the use of a relatively nearby sample of
tracers.  However, we also expect a large and distant data set to
constrain the gravitational potential at different positions, which is
ultimately a more reliable probe of the dark matter halo. The planned
or already-underway kinematically unbiased surveys, such as the
European Space Agency's astrometric mission Gaia (Turon et al. 2005),
the Space Interferometry Mission (SIM; Unwin et al. 2000), the RAdial
Velocity Experiment (RAVE; Steinmetz et al. 2006), the Sloan Digital
Sky Survey (SDSS; York et al. 2000), and the Sloan Extension for
Galactic Understanding and Exploration (SEGUE; Newberg et al. 2003,
Beers et al.  2004), will finally make it feasible to obtain the
required large sample of tracers with well-measured parameters.

Blue Horizontal-Branch (BHB) stars are excellent tracers of Galactic
halo dynamics, because they are luminous and have a nearly constant
absolute magnitude within a restricted color range (see, e.g., Sirko
et al. 2004 for BHB stars in SDSS). SDSS and SEGUE specifically
targeted BHB stars for spectroscopy; we employ these stars (and other
serendipitously discovered BHB stars with available spectra, e.g.,
mis-identified QSOs) to derive precision constraints on the mass of
the Milky Way's dark matter halo.

In \S 2 we describe the assembly of a particularly conservative (i.e.,
low contamination) selection of BHB stars from the spectra of
candidate BHBs in SDSS DR-6. Two different cosmological simulations,
one of a galaxy resembling the Milky Way with a $\sim 2 \times
10^{12}~{\rm M}_\sun $ halo, and one of a Milky-Way-like galaxy
simulation with a $\sim 1 \times 10^{12}~{\rm M}_\sun $ halo, are
presented in \S 3. This section also describes the estimation of the
circular velocity curve as a function of Galactocentric radius, $\rm
V_{cir}(r)$, obtained by comparing the observed BHB radial velocities
to the kinematics of the simulations. In \S 4 we present the resulting
estimates of the Milky Way's halo circular velocity curve, and discuss
implications for the virial mass of the Milky Way's halo. Our
conclusions are summarized in \S 5.

%%%%%%%%%%%%%%%%%%%%%%%%%%%%%%%%%%%%%%%%%%%%%%%%%%%%%%%%%%%%%%%%%%%%%%%%%%%%%%%%%%%%

\section{Data}

The SDSS is an imaging and spectroscopic survey covering more than a
quarter of the sky (York et al. 2000). Although its main focus was
(and is) extragalactic science, there are still large numbers of stars
not only imaged, but also targeted spectroscopically. In 2005 the
project entered a new phase (SDSS-II), in which SEGUE, a sub-survey of
SDSS-II, addresses fundamental questions about the formation and
evolution of our own Galaxy (http://www.sdss.org). In addition to
further (low-latitude) imaging, SEGUE in particular provides for a
more systematic and extensive acquisition of spectroscopy for stars in
the Milky Way, from which stellar parameters and radial velocities can
be derived from the application of a well-calibrated and well-tested
set of procedures, the SEGUE Stellar Parameter Pipeline (SSPP; see Lee
et al. 2008a,b; Allende Prieto et al.  2008).

The Sloan Digital Sky Survey uses a CCD camera (Gunn et al. 1998) on a
dedicated 2.5m telescope (Gunn et al. 2006) at Apache Point
Observatory, New Mexico, to obtain images in five broad optical bands
($ugriz$; Fukugita et al.~1996). The survey data-processing software
measures the properties of each detected object in the imaging data in
all five bands, and determines and applies both astrometric and
photometric calibrations (Lupton et al. 2001; Pier et al. 2003;
Ivezi\'c et al.~2004). Photometric calibration is provided by
simultaneous observations with a 20-inch telescope at the same site
(Hogg et al.~2001; Smith et al.~2002; Stoughton et al.~2002; Tucker et
al.~2006). This work is based on stellar spectra that are part of the
SDSS DR-6 (Adelman-McCarthy et al. 2008).

The radial velocities are taken from the SSPP, which (primarily) uses
matches of the observed spectra to a set of $908$ ELODIE template
spectra, corrected to the heliocentric standard-of-rest (HSR) frame.
Note that the SSPP automatically corrects for a systematic offset in
SDSS stellar radial velocities of 7.3 km s$^{-1}$ (see Lee et al.
2008a,b for derivation of this offset). After this correction is
applied, the SDSS radial velocities exhibit negligible systematic
errors, and have an accuracy between 5 and 20 km $\rm s^{-1}$,
depending on the S/N of the spectrum and the stellar spectral type.

\subsection{Sample Selection}

We aim to select as ``pure'' a set of true BHB stars as possible,
where the contamination, e.g., from halo blue-straggler stars, is
minimized, even if this selection procedure results in a smaller
sample. We have not made any effort to construct a complete sample of
BHB stars. Therefore, we have not simply followed the DR-6
classification procedures, but have employed a very stringent approach
combining previously established color cuts with a set of Balmer-line
profile selection criteria. Our technique is similar to that of Sirko
et al. (2004), but we use slightly different criteria in the adopted
color cuts. Throughout this paper all magnitudes are corrected for
Galactic extinction, and the colors are corrected for reddening, both
based on the procedures of Schlegel et al. (1998).

\subsubsection{Color Cuts}

\begin{figure}
\includegraphics[width=\textwidth]{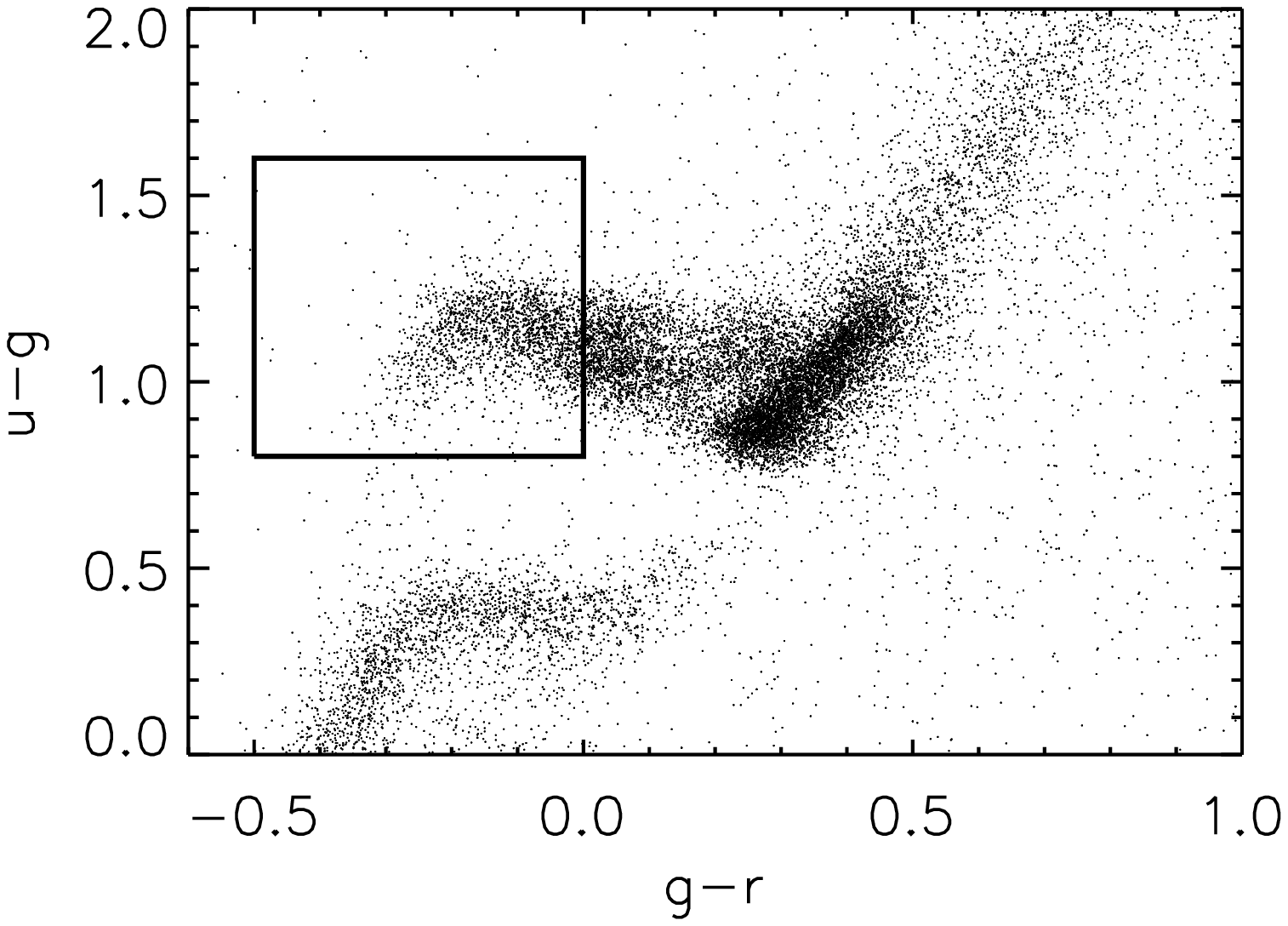}
\caption{SDSS color-color diagram showing all spectroscopically
  targeted objects that were subsequently confirmed as stars. The
  large Balmer jump of A-type stars places them in the region,
  where our ``color cut'' selection box is drawn. This color selection
  approach follows Yanny et al. (2000).}
\label{f:f1}
\end{figure}

We start our selection of BHB stars by adopting the color cuts for
identification of BHB candidates used in Yanny et al. (2000):

%% for bright stars ($ g < 18 $)
\begin{displaymath} \ \rm 0.8<u-g<1.6
\end{displaymath}
\begin{displaymath} \ \rm -0.5<g-r<0.0
\end{displaymath}
\noindent These color cuts are shown in Figure~\ref{f:f1}; the
rectangle is the region of BHB candidates. This color cut produces
$\sim 10,000$ BHB photometric candidates\footnote{A data file of the
  full set of candidates is made available in electronic form in the
  online edition of the ApJ, see Table 1 for example.} with existing
spectra, but with a considerable contamination by both blue straggler
(BS) stars and warm main-sequence stars (MS). The subsequent
spectroscopic analysis for BHB candidates is aimed at eliminating, or
at least greatly reducing, contamination from such stars.

\subsubsection{Balmer-Line Profile Cuts}

\begin{figure}
\includegraphics[width = 16cm,height=10cm]{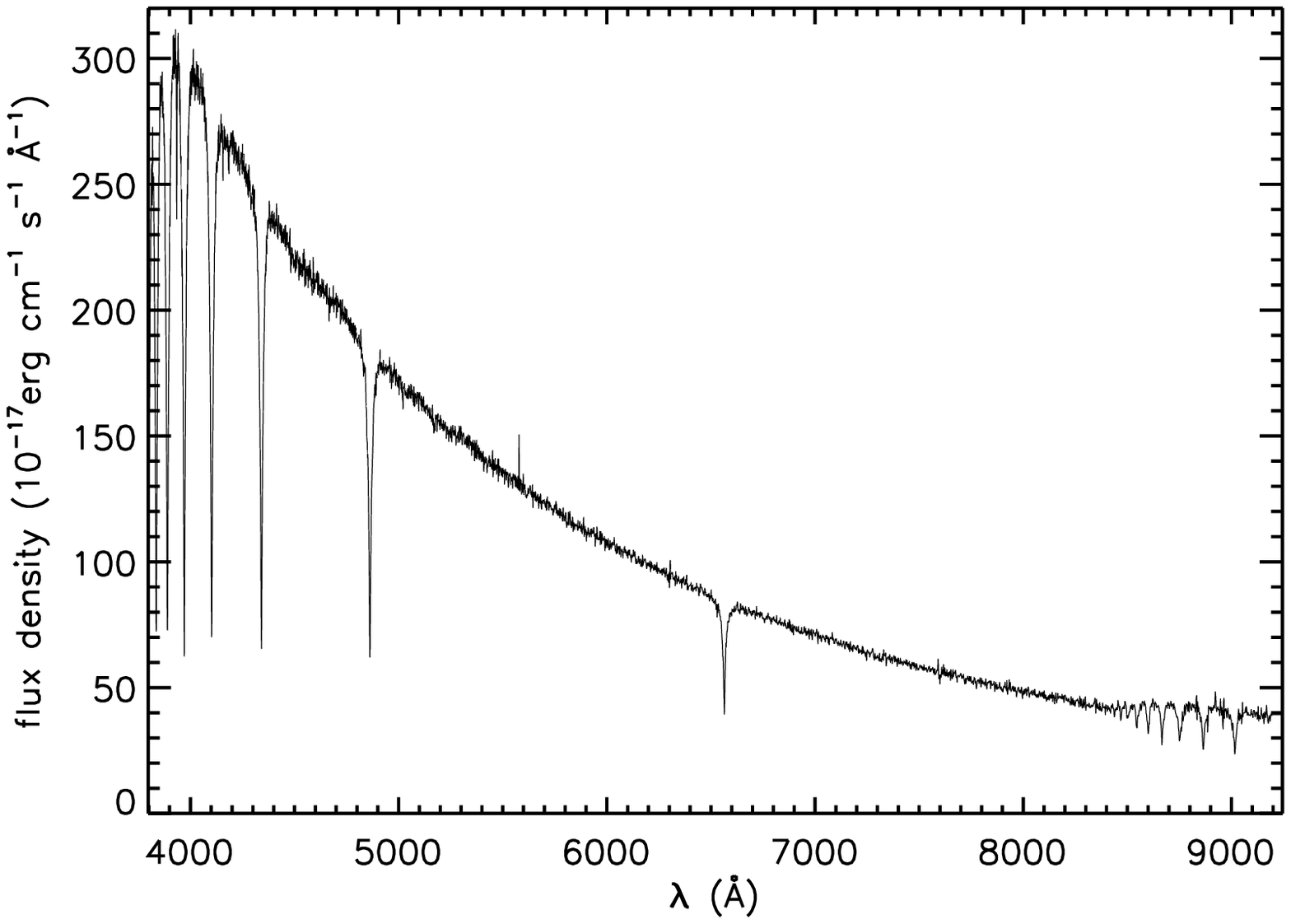}
\includegraphics[width = 16cm,height=10cm]{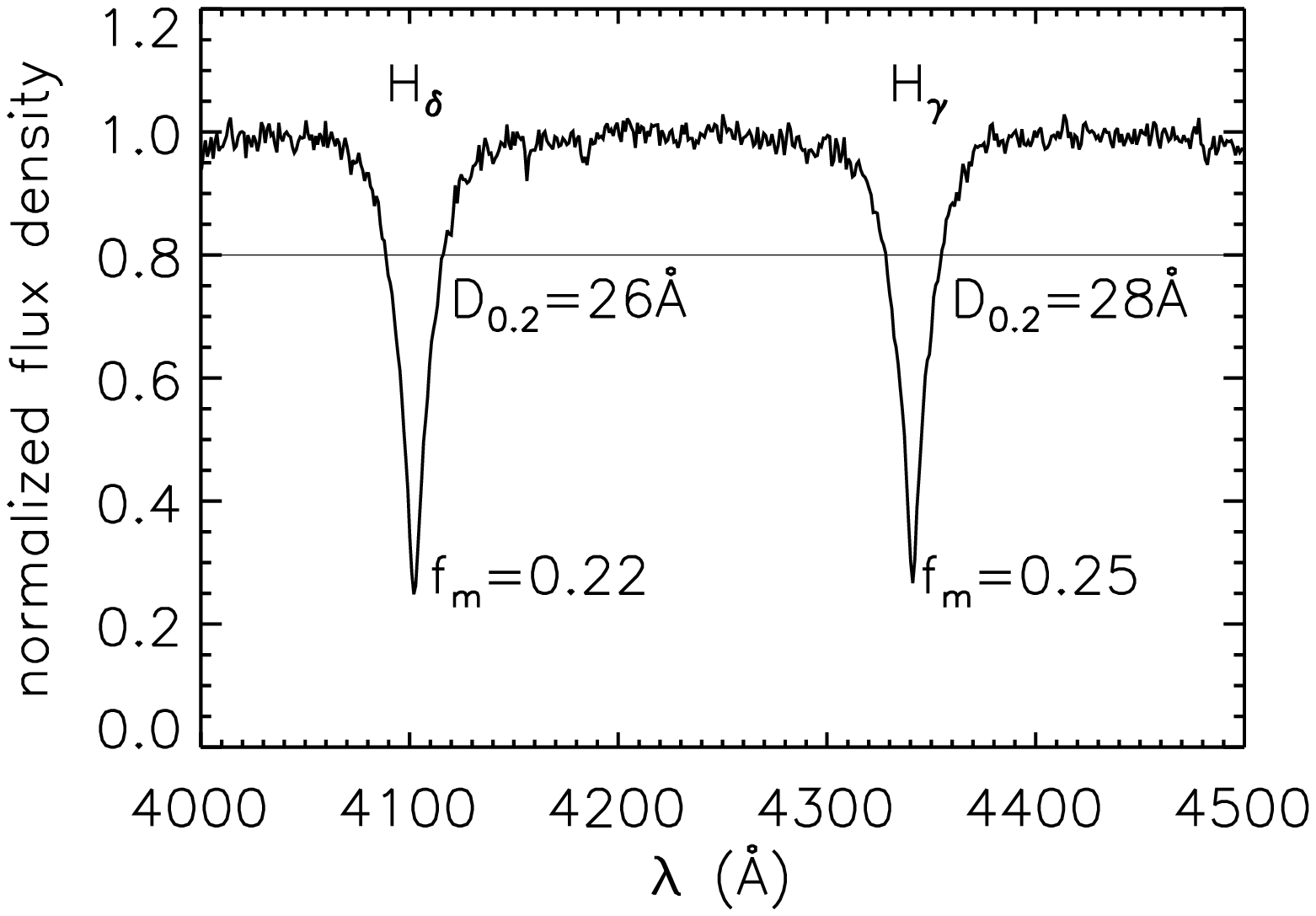}
\caption{Spectrum of a typical (high-S/N) BHB star (upper panel), and
  the $\rm H_\gamma ~\sim~ H_\delta$ region of the same star with the
  continuum divided out (lower panel). The $\rm (f_m,D_{0.2})$
  parameters that are used for the sample selection are labelled for
  both lines (see \S 2.1.2 for discussion).}
\label{f:f2}
\end{figure}

In the ranges of effective temperature considered herein (roughly
7000~K to 10,000~K), BHB stars have lower surface gravities than BS
stars and higher temperatures than (old, halo population) MS stars.
The Balmer-line profiles of warm stars are sensitive to both gravity
and temperature; their analysis provides a powerful method to select
BHB stars with confidence. We analyze the line profiles after
normalizing the continuum for all stars, as illustrated in
Figure~\ref{f:f2}. Then we combine two independent methods to identify
non-BHB stars, as described below.

\begin{figure}
\includegraphics[width=\textwidth]{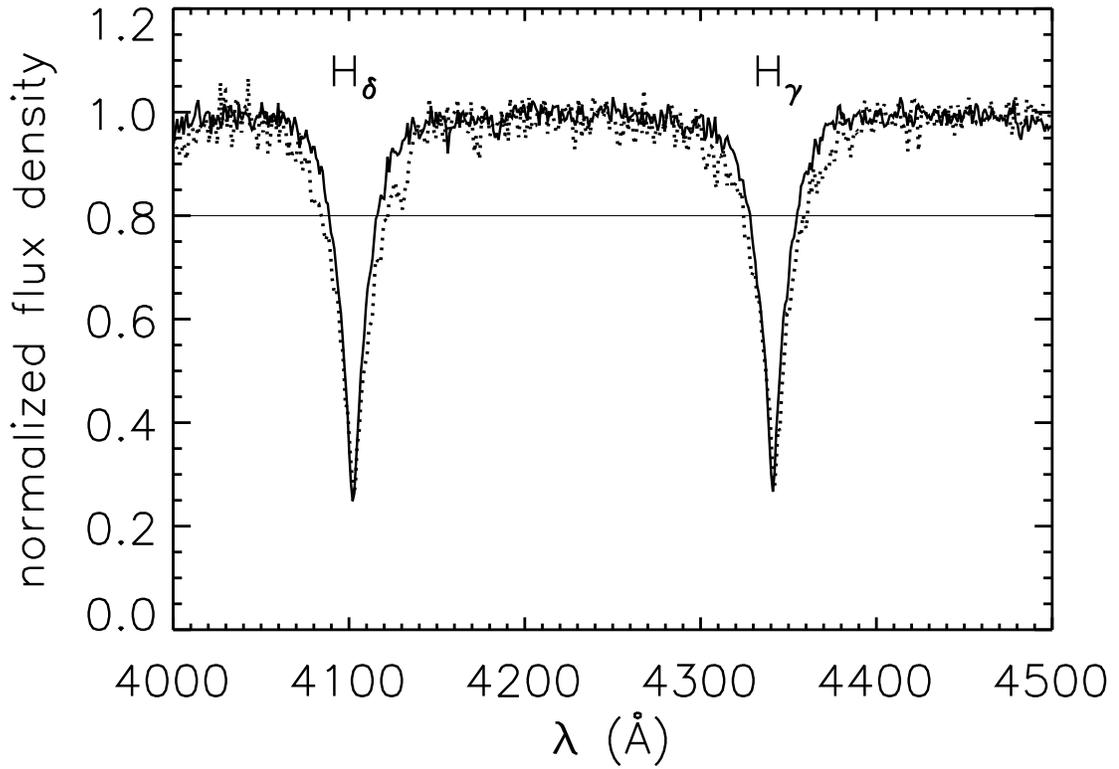}
\caption{Normalized spectrum of a BHB star (solid line) and a BS star
  (dotted line), of similar effective temperatures, in the $\rm
  H_\gamma ~\sim~ H_\delta$ region. Although subtle, one can notice
  that the BS stars's Balmer lines are wider at $20\%$ below the local
  continuum than the BHB star. These effects arise due to the higher
  gravity of the BS star.}
\label{f:f3}
\end{figure}

\begin{figure}
\includegraphics[width=\textwidth]{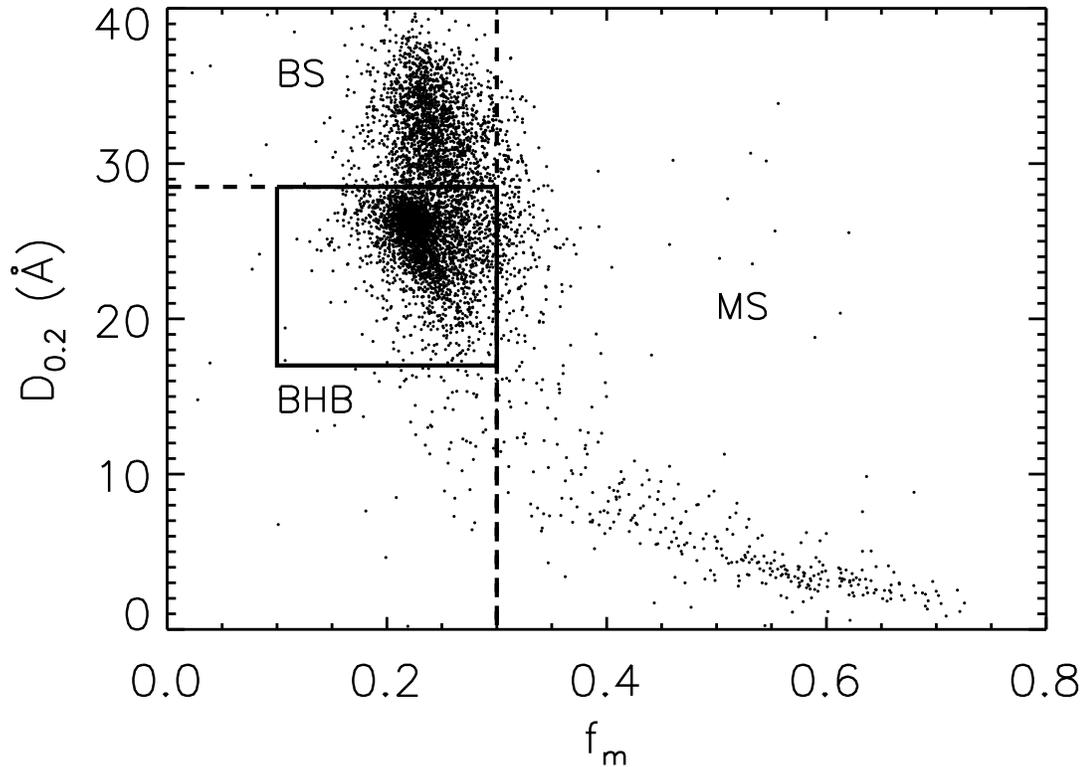}
\caption{The parameters $\rm f_m$ and $\rm D_{0.2}$, as determined
  from the $\rm H_\delta$ line, for stars brighter than $\rm g~=~18$
  and passing the color cuts shown in Figure~\ref{f:f1}. The trail of
  stars with $\rm f_m \geqslant 0.30$ are too cool to be BHB stars,
  while the concentration of stars with $\rm D_{0.2} \geqslant
  28.5~\rm \AA$ is due to blue stragglers with higher surface gravity.
  The stars that lie well outside the main locus are the result of
  poor parameter determinations due to missing spectroscopic data at
  the location of the $\rm H_\delta$ line. The region enclosed by the
  box is used as the BHB selection criterion for the $\rm H_\delta$,
  $D_{0.2}$ \& $f_m$ method.}
\label{f:f4}
\end{figure}

The $D_{0.2}\ vs.\ f_m$ method (Pier 1983; Sommer-Larsen, Christensen
1986; Arnold \& Gilmore 1992; Flynn, Sommer-Larsen, \& Christensen
1994; Kinman, Suntzeff, \& Kraft 1994; Wilhelm, Beers, \& Gray 1999)
discriminates BHB stars from BS stars by determining the value of $\rm
D_{0.2}$, the width of the Balmer line at 20\% below the local
continuum, and distinguishes BHB stars from MS stars by measuring the
value of $\rm f_m$, the flux relative to the continuum at the line
core (Beers et al. 1992; Sirko et al. 2004). Figure~\ref{f:f3} shows
how $\rm D_{0.2}$ can distinguish a BHB star from a BS star. A plot of
$\rm D_{0.2}$ versus $\rm f_m$ of the $\rm H_\delta$ line for stars
brighter than $g~=~18$, and passing the initial color cuts, is shown
in Figure~\ref{f:f4}. The concentration of stars centered at $\rm
(f_m, D_{0.2}) = (0.23, 25\rm \AA$) represents the BHB stars; the
stars with larger $\rm D_{0.2}$ are BS stars, and the remaining stars
are MS stars. Figure~\ref{f:f4} indicates that the sample
contamination resulting from the color-selection only is rather
severe, on the order of 50\%. The criteria for confirmation of a BHB
star, based on the H$\rm _\delta$ line analysis, are:

\begin{displaymath}
\ 17~\rm \AA \leqslant D_{0.2}\leqslant 28.5 ~\rm \AA,    \ 0.1 \leqslant
f_m \leqslant 0.3\ .
\end{displaymath}

The {\it scale width vs. shape} method (Clewley et al. 2002) is based
on a S\'{e}rsic profile (S\'{e}rsic 1968) fit to the Balmer lines:
\begin{equation}
  \rm y = 1.0 - a\rm \exp{\left[-\left(\frac{|\lambda-\lambda_0|}{b}\right)^c\right]},
\end{equation} 

\noindent where $\rm y$ is the normalized flux density and $\rm
\lambda_0$ is assumed to be the nominal wavelength of the Balmer line.
To account for small radial velocity corrections and the imperfect
normalization of spectra, we fit the normalized extracted spectrum to
the S\'ersic profile with five free parameters : $\rm a, b$, $\rm c$,
$\rm \lambda_0$ and $\rm n$.

\begin{equation}
\rm y =  n - a 
\exp{\left[-\left(\frac{|\lambda-\lambda_0|}{b}\right)^c\right]}
\end{equation}

\begin{figure}
\includegraphics[width=\textwidth]{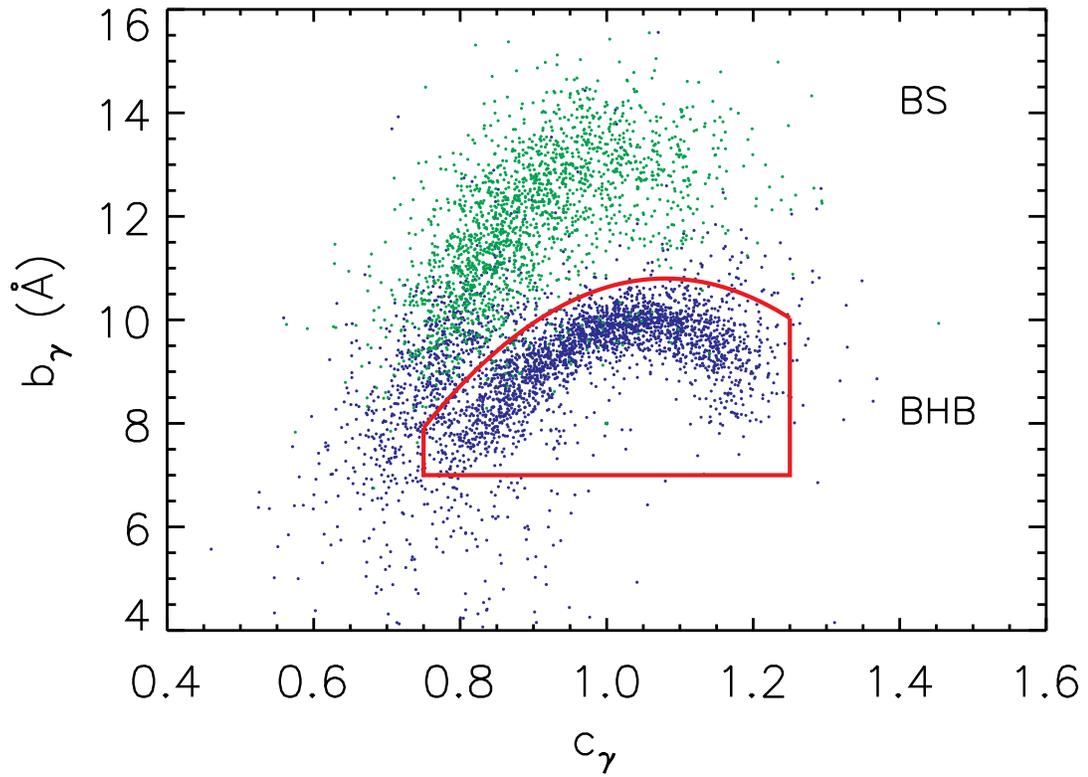}
\caption{Green dots are BS stars and blue dots are BHB stars
  identified by the $\rm D_{0.2}$ \& $\rm f_m$ methods. They can be
  distinguished clearly by the ``gap". The closed region indicates the
  $\rm H_\gamma$ scale width-shape criteria that selects BHB stars.}
\label{f:f5}
\end{figure}

The set of stars that passed the initial color cuts exhibits a bimodal
distribution in the $\rm c_\gamma$ versus $\rm b_\gamma$ plane
(Figure~\ref{f:f5}), where $\gamma$ refers to the H$_\gamma$ line.
This gap allows one to quite cleanly separate BHB stars from BS stars,
according to:

\begin{displaymath} \rm 0.75 \leqslant c_\gamma \leqslant 1.25
\end{displaymath}
\begin{displaymath} \rm 7.5~\rm \AA \leqslant b_\gamma \leqslant
  10.8-26.5\left(c_\gamma-1.08\right)^2\ .
\end{displaymath}

As the color coding of the points in Figure~\ref{f:f5} shows, most BHB
stars selected by the $D_{0.2}$ \& $f_m$ method already lie in the
appropriate region of the {\it scale width vs. shape} method applied
to $\rm H_\gamma$. The combination of these two stringent criteria
indeed appears to eliminate most stars that are not {\it bona-fide}
BHB stars. The combination of Figure~\ref{f:f4} and Figure~\ref{f:f5}
suggests that the contamination is well below 10\%.

Note that the two spectroscopic criteria are determined based on the
spectra of bright stars ($g\leqslant18$). Because of the high quality
spectra available for such stars, the criteria can be identified
easily by eye from Figure~\ref{f:f4} and Figure~\ref{f:f5}. For
fainter stars ($g>18$) we adopt the same spectroscopic criteria as for
the bright stars.

There are a total of $2558$ stars\footnote{A data file of the full set
  of adopted BHB stars is made available in electronic form in the
  online edition of the ApJ, see Table 2 for example.} that survive
the color cuts, and {\it both} of the Balmer-line profile cuts
described above. This sample forms the basis of our remaining
analysis.

\subsection{The Absolute Magnitude of BHB Stars}

\begin{figure}
\includegraphics[width=\textwidth]{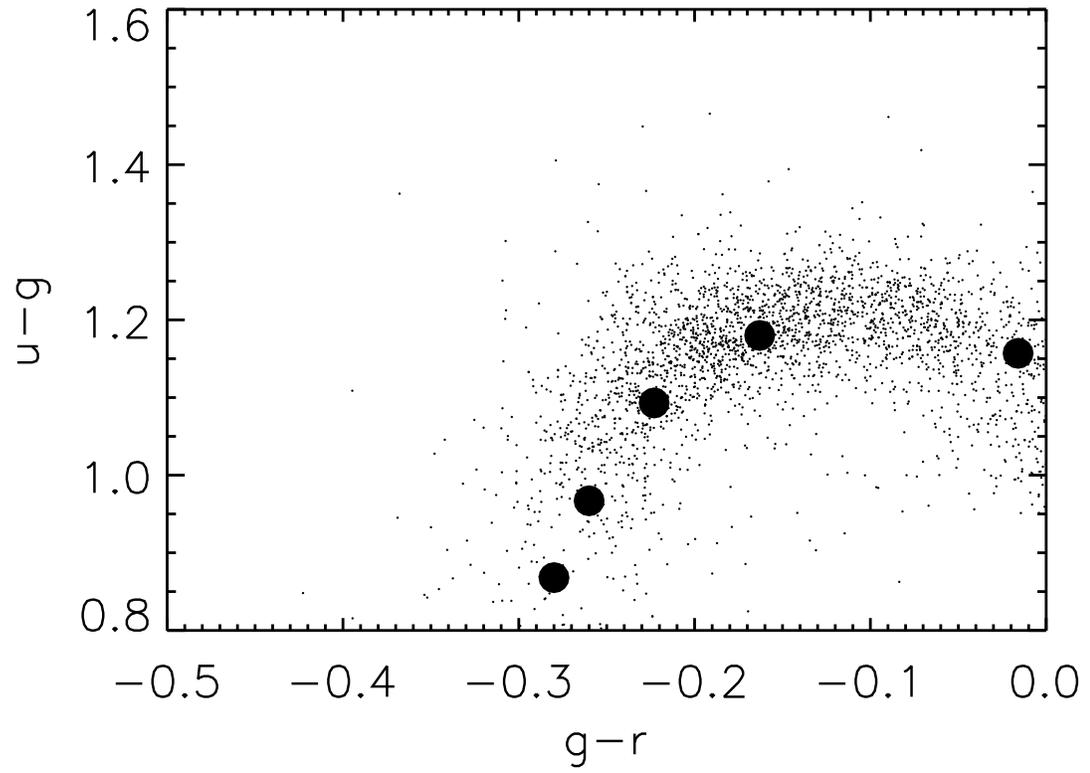}
\caption{Color-color ($\rm u-g$ vs. $\rm g-r$) diagram of our BHB
  stars. The five black dots, starting from the right to the left,
  represent the model colors for BHB stars of absolute magnitudes $\rm
  M_{g} = 0.60, 0.55, 0.65, 0.70, 0.80$.}
\label{f:f6}
\end{figure}

Our basic approach is to identify BHB stars from their spectra (\S
2.1), then estimate their absolute magnitude from photometry alone.
BHB stars have similar, but not identical, absolute magnitudes, as
they are affected slightly by temperature and metallicity (e.g.,
Wilhelm, Beers, \& Gray 1999; Sirko et al.  2004). Figure~\ref{f:f6}
shows five theoretical absolute magnitudes $\rm M_{g} = 0.60, 0.55,
0.65, 0.70, 0.80$ in the $\rm (u-g, g-r)$ plane, taken from Sirko et
al. (2004). For each BHB star in our sample we define the most
probable absolute magnitude associated with its $\rm (u-g, g-r)$
colors by simply finding the absolute magnitude of the point on the
theoretical track that is closest to the observed star in this
color-color space.\footnote{The method described above could be
  improved upon by fitting a polynomial to the theoretical absolute
  magnitudes and color index ($u-g,g-r$), and use it instead of
  picking the ``closest'' model. However, the difference between the
  two methods is less than the theoretical uncertainties in the
  derived luminosities.} The absolute magnitude error of a given star
derived from this method is on the order of 0.2 mag (Sirko et al.
2004), corresponding to a distance accuracy of 10\%; errors in the
measured photometry are much smaller than this error. The distance
from the Sun, and from the Galactic center can then be determined
from:

\begin{equation} g={\rm M}_g+5\log_{10}d-5
\end{equation}
\begin{eqnarray} r^2=({\rm R}_\sun-d\cos b\cos l)^2+d^2\sin^2
  b+d^2\cos^2 b\sin^2 l,
\end{eqnarray}

\noindent where $g$ is the extinction-corrected magnitude in the $g$
band, ${\rm M}_g$ is the absolute magnitude in the $g$ band, $ d $ is
the distance to the Sun, $ r$ is the distance from the Galactic
center, $ b$ and $ l$ are the Galactic latitude and longitude
respectively; we take ${\rm R}_\sun$, the distance of the Sun from the
Galactic center, to be $8.0$~kpc.

\subsection{The Spatial, Velocity and Metallicity Distribution of the
  BHB Star Sample}

For ease of the subsequent analysis, we convert the heliocentric
radial velocities to the Galactic standard of rest (GSR) frame by
adopting a value of $220$ km s$^{-1}$ for the Local Standard of Rest
(V$_{\rm lsr}$) and a Solar motion of $(+10.0, ~+5.2, ~+7.2)$ km
s$^{-1}$ in (U,V,W), which are defined in a right-handed Galactic
system with U pointing towards the Galactic center, V in the direction
of rotation, and W towards the north Galactic pole (Dehnen \& Binney
1998). Hereafter, $\rm V_{los}$ stands for the radial velocity in the
GSR frame (i.e., the radial velocity component along the star-Sun
direction, corrected for Galactic rotation). If $\rm V_{helio}$ is the
heliocentric radial velocity, then:

\begin{eqnarray} {\rm V}_{\rm los}~=~{\rm V}_{\rm helio}~+~10.0 ~{\rm km~s}^{-1}~\times~\cos l\cos b+\nonumber \\
  7.2~{\rm km~s^{-1}}~\times~\sin b+({\rm V_{lsr}}+5.2)~{\rm
    km~s}^{-1}~\times~\sin l\cos b.
\end{eqnarray}

Our sample of $2558$ BHB stars may contain some thick-disk stars, with
$1~\rm kpc<|z|<4~\rm kpc$, so we impose an additional geometric
constraint $\rm |z|>4~ kpc$, which reduces the sample to $2401$
(presumed) {\it halo} BHB stars within 60 kpc, and with radial
velocity error less than 30 km $\rm s^{-1}$.

\begin{figure}
\includegraphics[width=16cm,height=10cm]{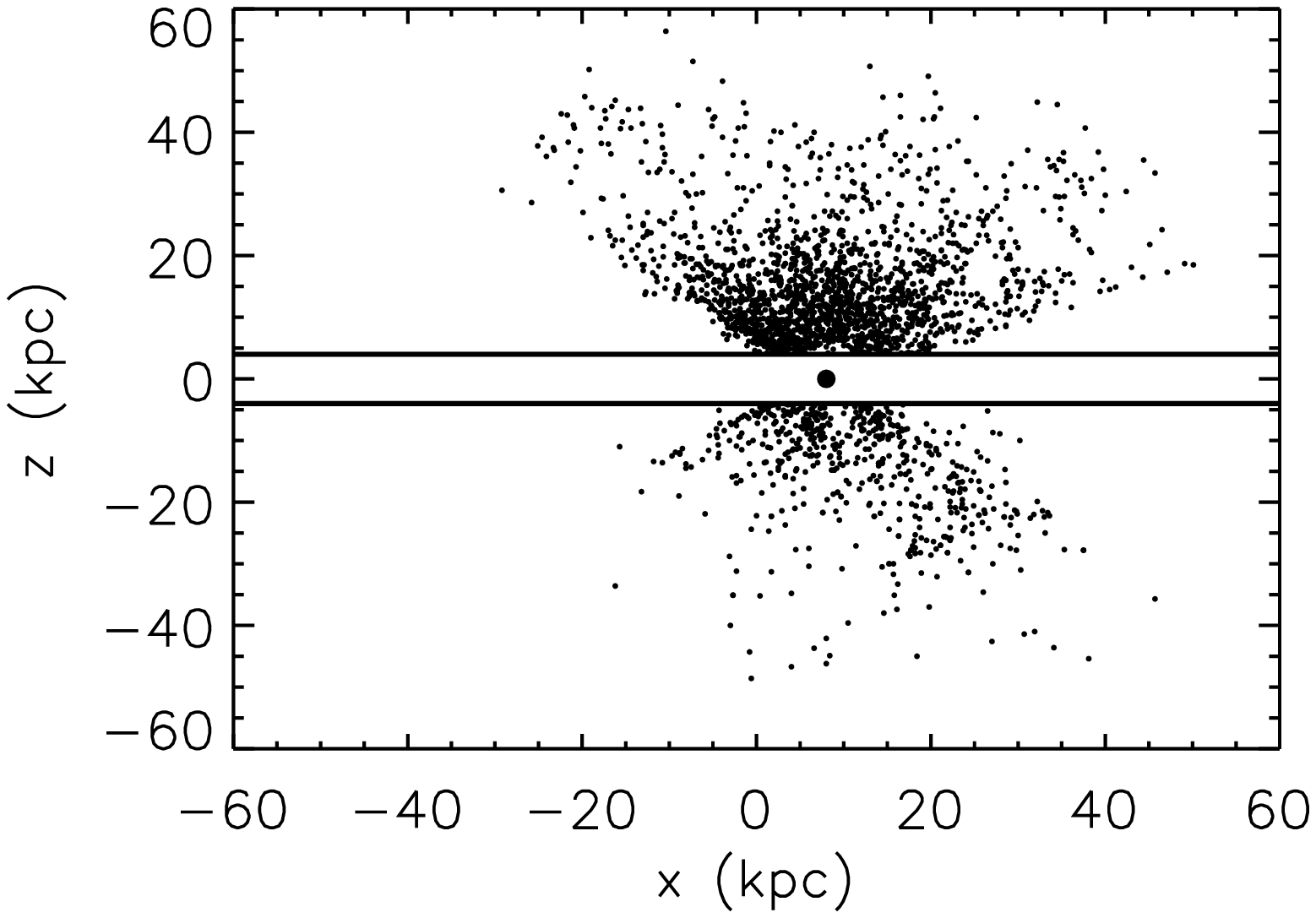}
\includegraphics[width=16cm,height=10cm]{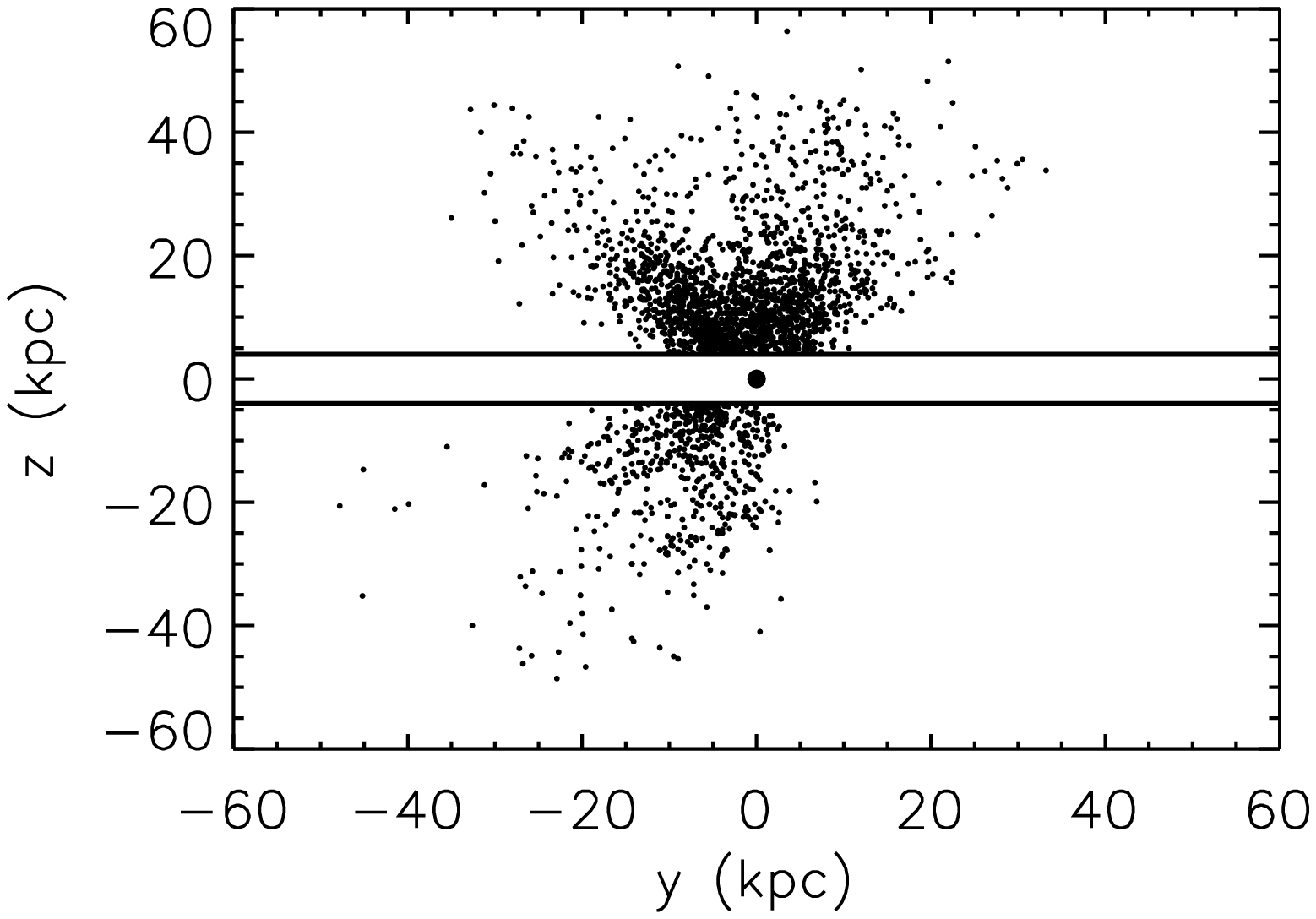}
\caption{The spatial distribution of the BHB stars in our sample,
  projected in the z-x (upper panel) and z-y plane (lower panel),
  respectively. The large filled circle is the Sun; the two lines are
  the planes 4 kpc above and below the Galactic disk plane.}
\label{f:f7}
\end{figure}

\begin{figure}
\includegraphics[width=\textwidth,height=10cm]{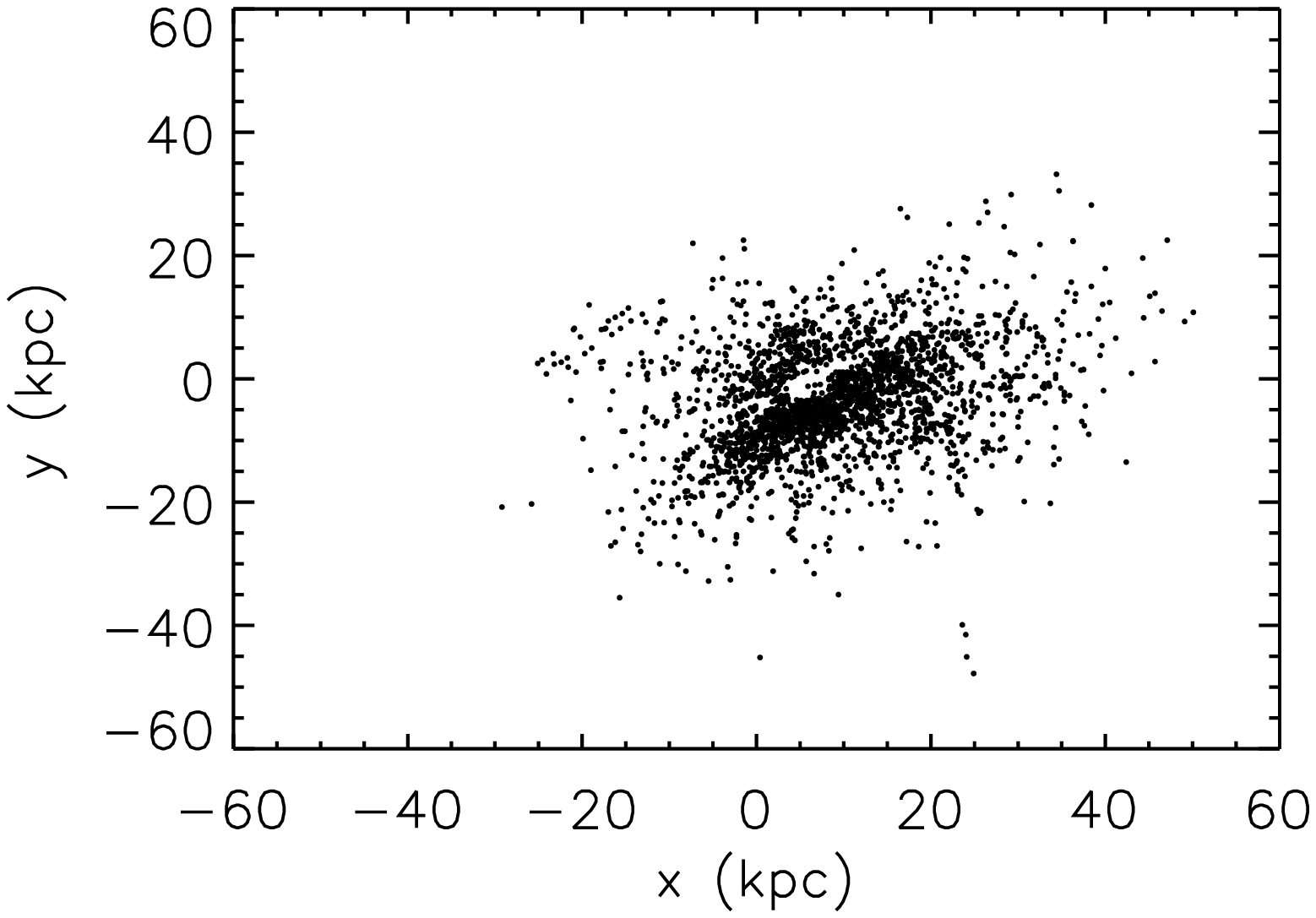}
\includegraphics[width=\textwidth,height=10cm]{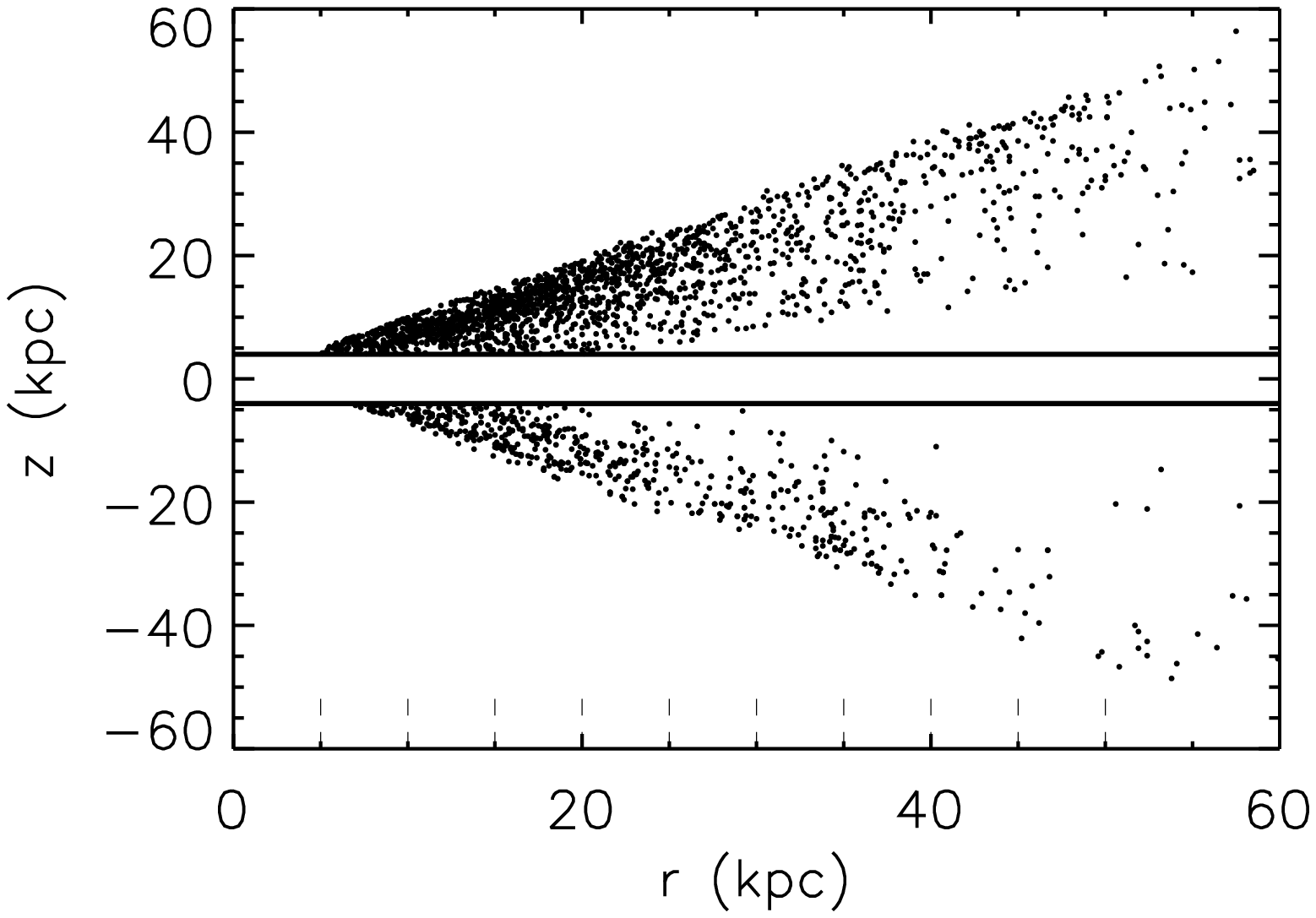}
\caption{The spatial distribution of the sample BHB stars in the x-y
  plane (upper panel) and z-r plane (lower panel). The coordinate
  origin is the location of the Galactic center. The two lines in the
  z-r plot are the planes 4 kpc above and below the Galactic disk
  plane. The short dashed vertical lines in the lower panel show the
  bin boundaries chosen for the subsequent analysis.}
\label{f:f8}
\end{figure}

\begin{figure}
\includegraphics[width=\textwidth,height=10cm]{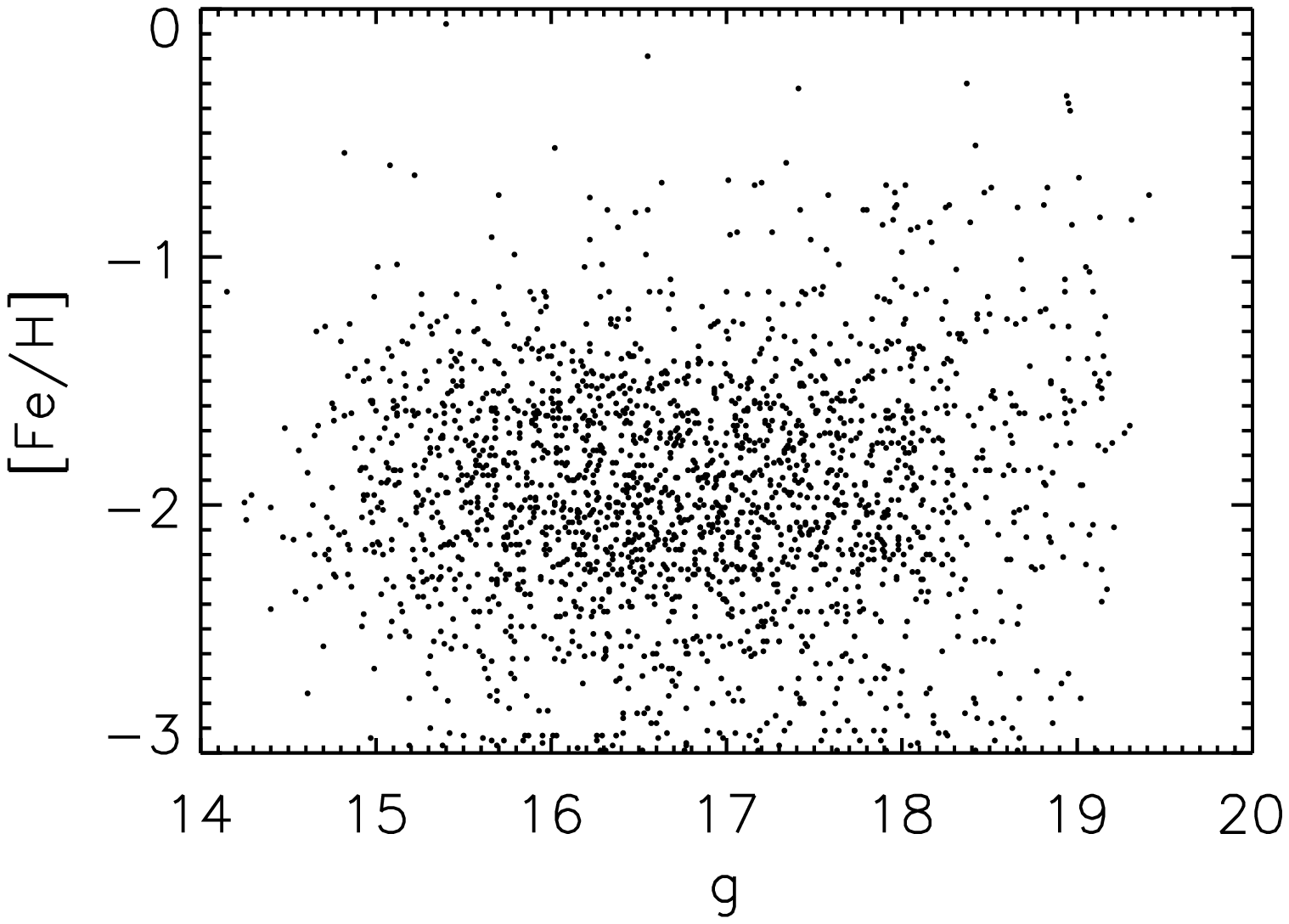}
\includegraphics[width=\textwidth,height=10cm]{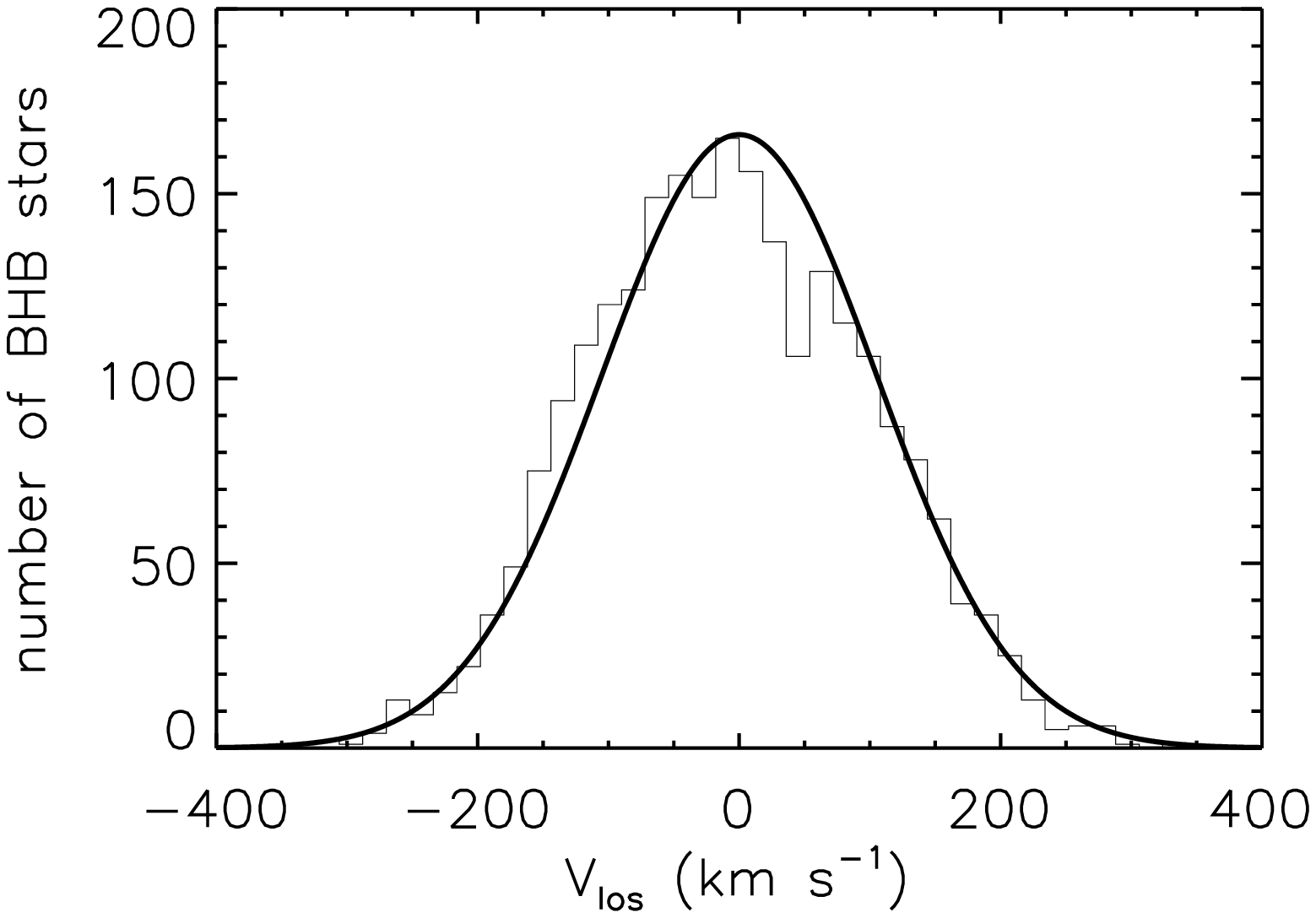}
\caption{(Upper panel) The distribution of metallicities, [Fe/H], as a
  function of apparent magnitude, for the entire sample of halo BHB
  stars. (Lower panel) The distribution of line-of-sight velocities,
  corrected to the GSR, for the entire sample of BHB stars. A Gaussian
  of width $\sigma = 105~$km $\rm s^{-1}$ centered on the local
  standard of rest is shown for reference. }
\label{f:f9}
\end{figure}

\begin{figure}
\includegraphics[width=\textwidth,height=10cm]{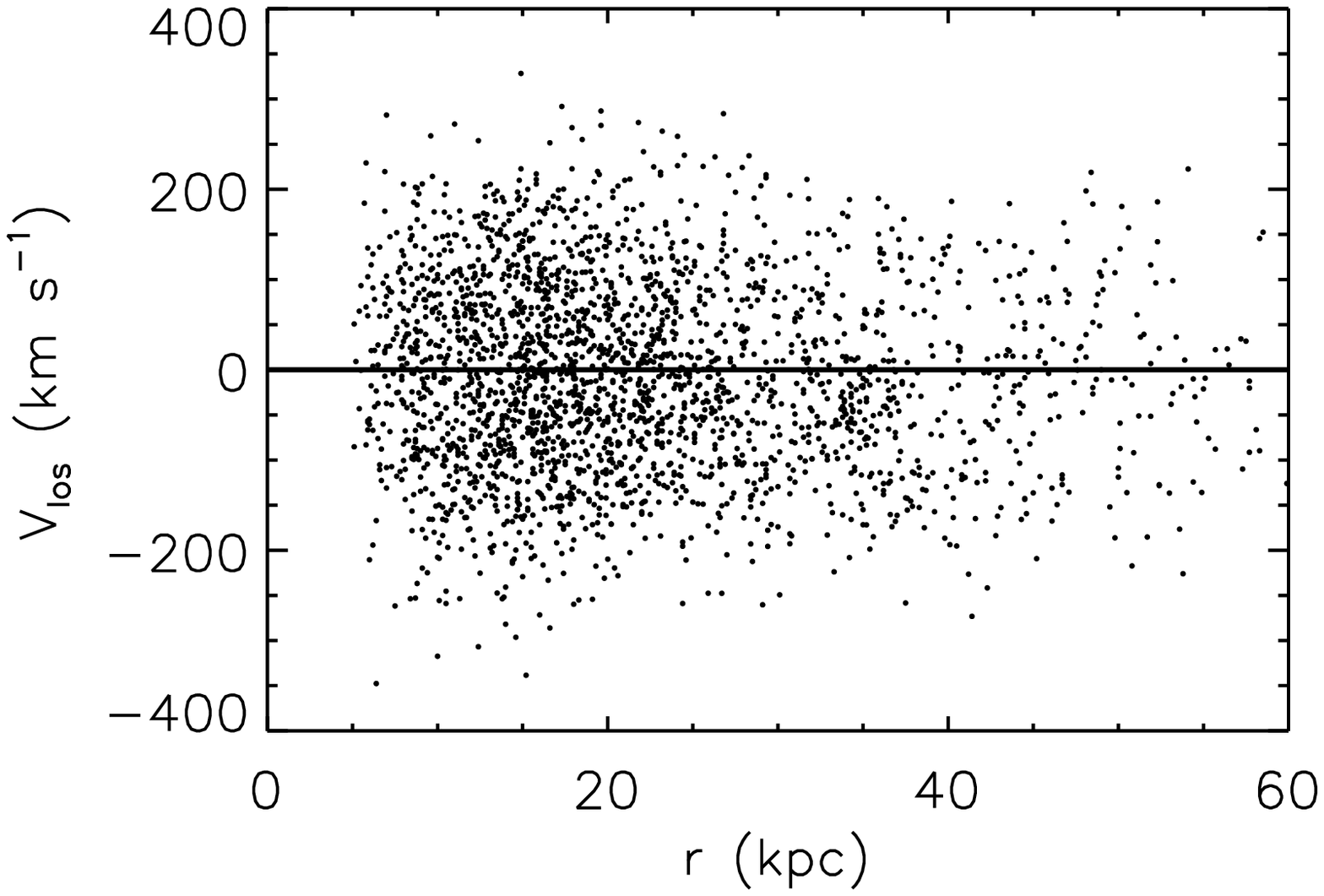}
\includegraphics[width=\textwidth,height=10cm]{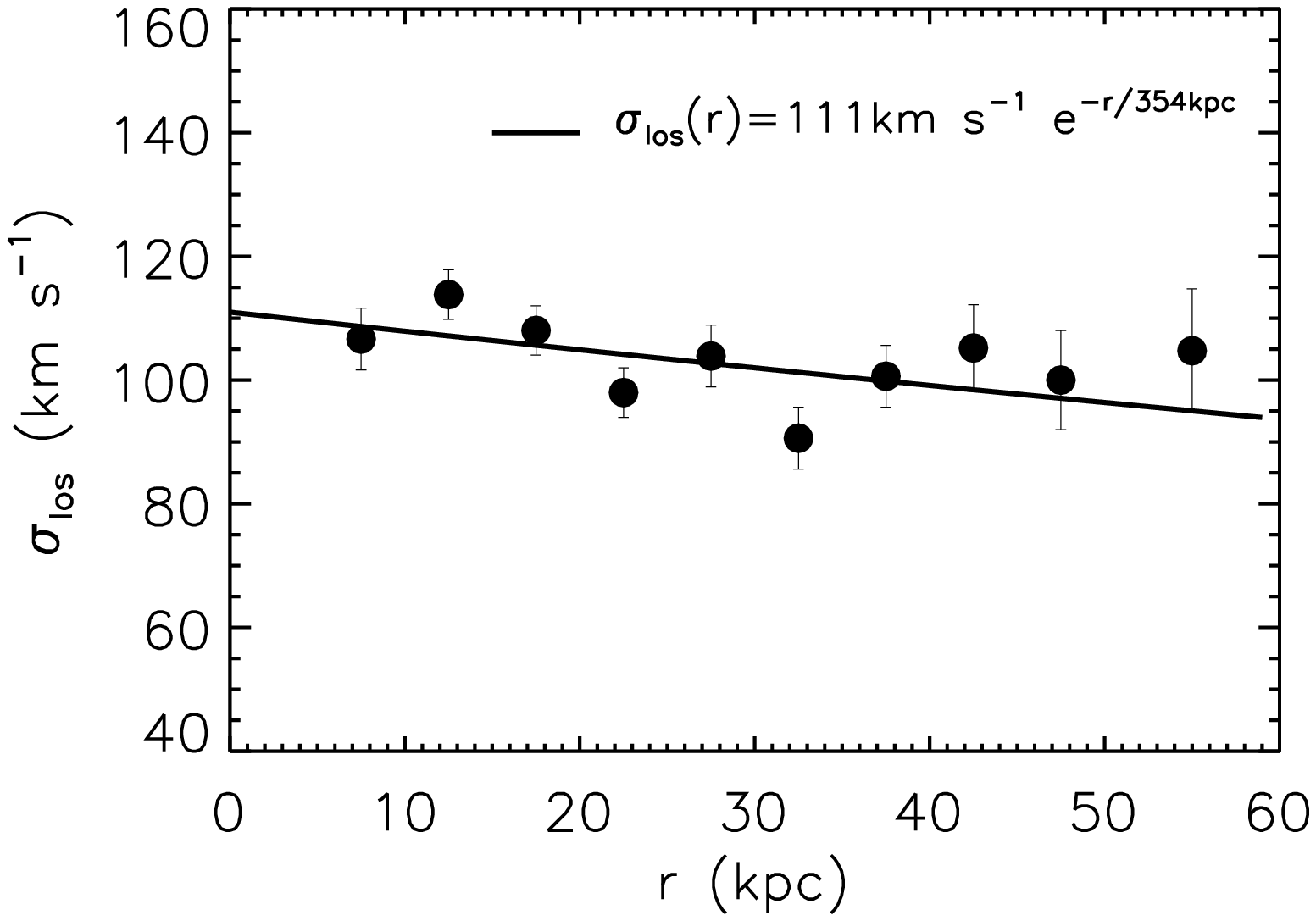}
\caption{(Upper panel) The distribution of $\rm V_{los}$ as a function
  of Galactocentric distance, $\rm r$, for the entire sample of halo
  BHB stars. (Lower panel) The velocity dispersion, $\rm
  \sigma_{los}$, as a function of Galactocentric distance. A best fit
  exponentially falling relationship is plotted.}
\label{f:f10}
\end{figure}

The spatial distribution of our $2401$ halo BHB stars is shown in
Figure~\ref{f:f7} and Figure~\ref{f:f8}, with all stars located at
least $4$~kpc from the Galactic disk and at $5 - 60$~kpc from the
Galactic center. Figure~\ref{f:f9} (upper panel) shows the derived
[Fe/H] estimates for the BHB sample reported by the SSPP (see Lee et
al.  2008a,b for details).  Most of our BHB stars are metal-poor
([Fe/H] $\sim -$2), as expected for a sample of halo stars.  This is
further testament to the quality of the SDSS spectra and the rigor of
the sample selection.  The observed distribution of line-of-sight
velocities is well-fitted by a Gaussian distribution with $\rm
\sigma_{v_{los}} = 105 ~km~ s^{-1}$, as shown in the lower panel of
Figure~\ref{f:f9}.  The upper panel of Figure~\ref{f:f10} shows the
distribution of $\rm V_{los}$ vs. $\rm r$ for the {\it halo} BHB
stars. It reveals a nearly equal number of stars with positive and
negative $\rm V_{los}$ at a given radius: the BHB population exhibits
very little net rotation, so we will subsequently analyze only $\rm
V_{los}$.  The lower panel of Figure~\ref{f:f10} shows the binned
velocity dispersion, $\rm \sigma_{\rm los}$, of the sample as a
function of radius. It is well described by
\begin{equation}
\rm \sigma_{los}(r)={\rm      \sigma_0     \exp}
  \left(\frac{\rm -r}{\rm r_0}\right),~\rm  with ~ {\rm \sigma_0}=111^{+1}_{-1}~{\rm
    km~s}^{-1}~ and ~ \rm r_0~=~354^{+91}_{-60}~{\rm kpc}
\end{equation}

\noindent where r is the Galactocentric radius. We will use Eqn 6
subsequently for Jeans Equation modeling. A {\it maximum likelihood}
fit to estimate $\rm \sigma_0$ and $\rm r_0$ yields $\rm
\sigma_0=111^{+1}_{-1}~{\rm km~s}^{-1}~ and ~ \rm
r_0=427^{+240}_{-108}~{\rm kpc}$, which is consist with Eqn 6 within
1-$\sigma$. This indicates that the choice of the bins has small
effect on the inferred $\rm \sigma_ {los}(r)$. We also checked that
these values of $\rm \sigma_{los}(r)$ are insensitive to small changes
in the assumed $\rm V_{lsr}$.

\begin{figure}
\includegraphics[width=\textwidth,height=10cm]{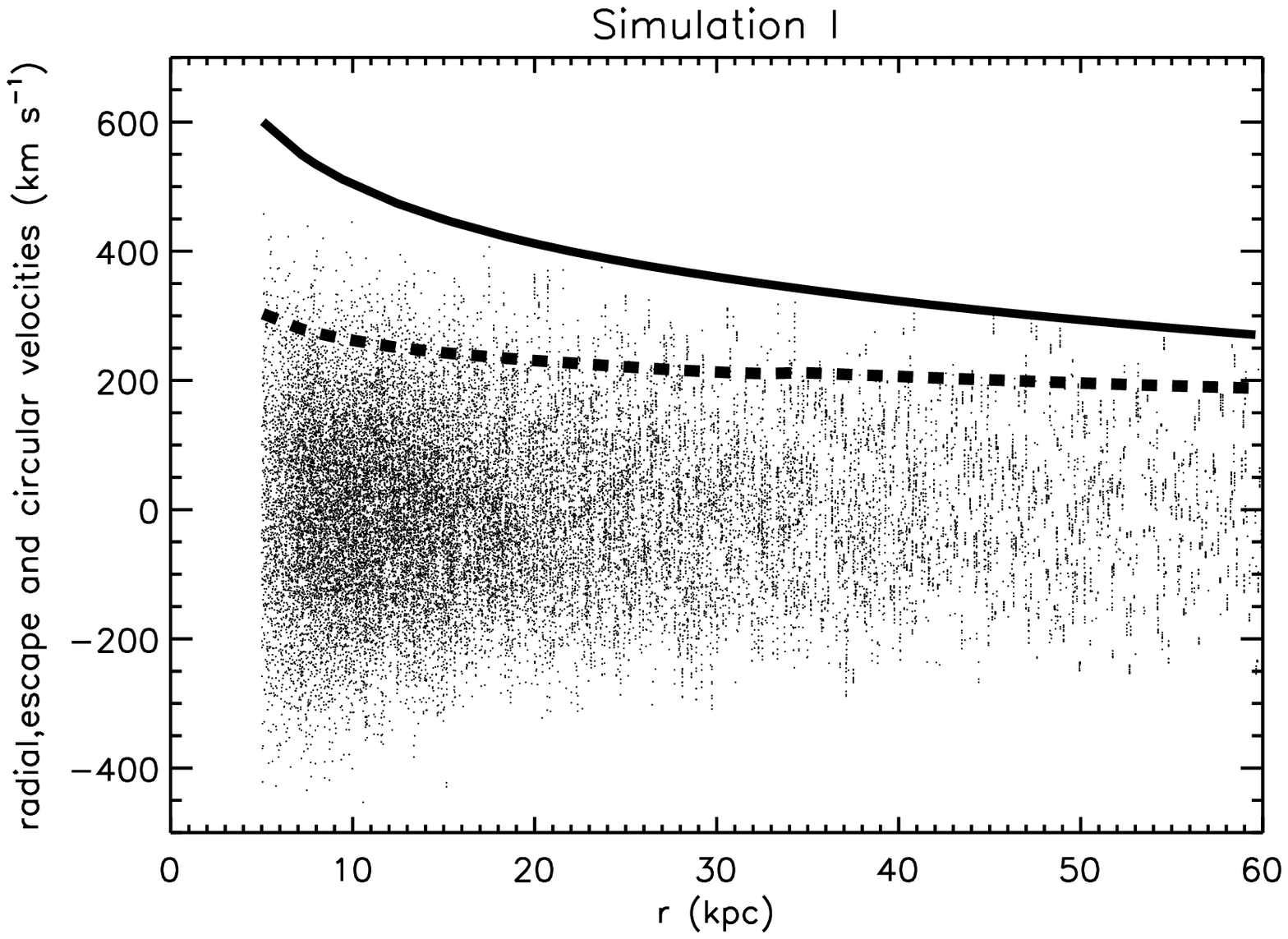}
\includegraphics[width=\textwidth,height=10cm]{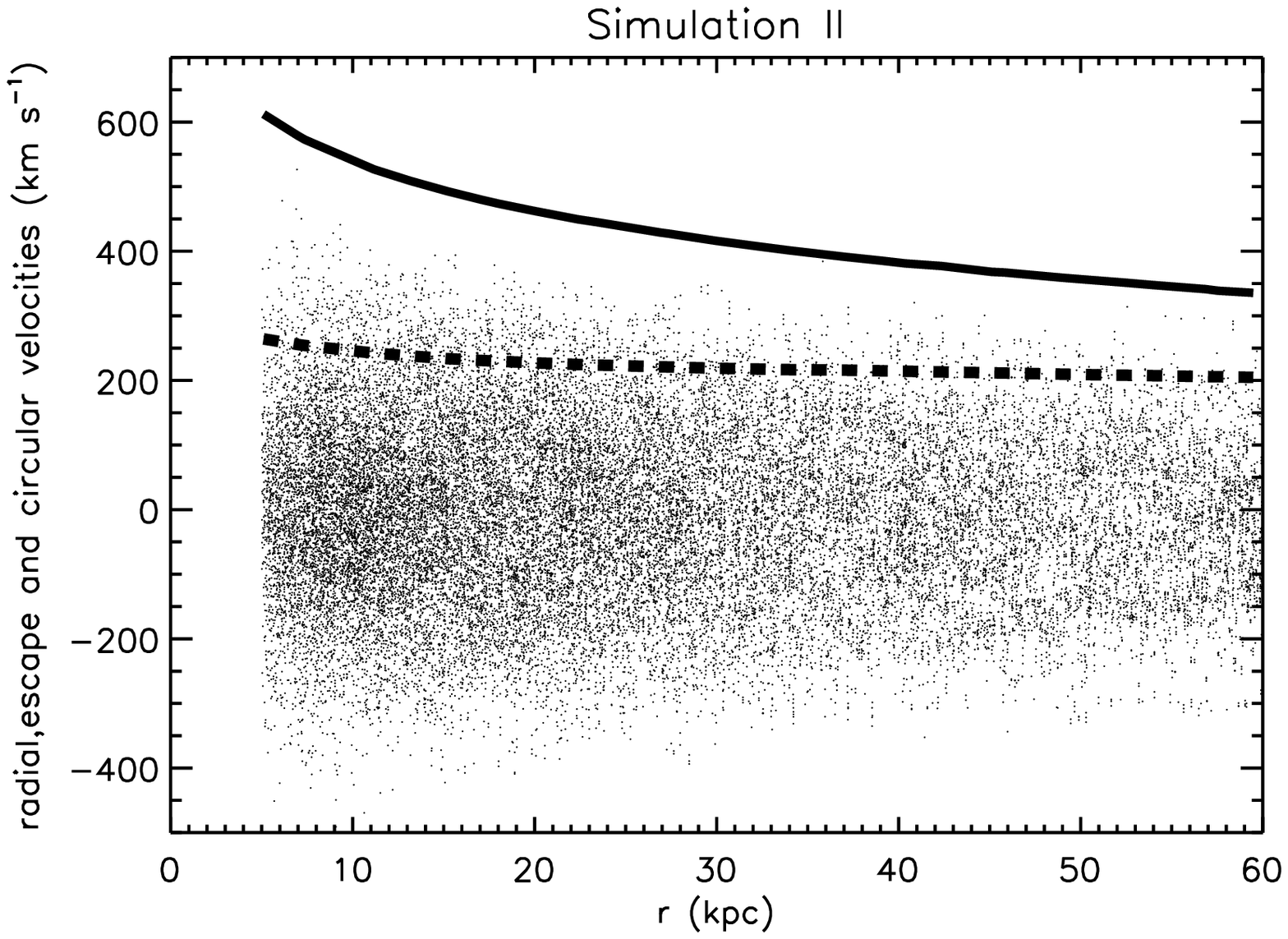}
\caption{The Galactocentric radial velocity, escape velocity and
  circular velocity distributions of the stars in Simulation I (upper
  panel) and in Simulation II (lower panel) (see \S 3); the
  simulations are ``viewed'' from the position of the Sun to lie
  within the SDSS DR-6 footprint. The solid line delineates the
  predicted escape velocity, while the dashed line indicates the
  predicted circular velocity. The dots represent the radial
  velocities.}

\label{f:f12}
\end{figure}

The radial number density profile of halo stars in the Milky Way in
the range $\sim 10-60$ kpc can be approximated by $\rm \rho\sim
r^{-3.5}$ (e.g., Bell et al 2007). Accounting for the $\rm r^2{\rm
  d}r$ volume effect, inspection of Figure~\ref{f:f10} and
Figure~\ref{f:f12} reveals that the radial distribution of our BHB
sample falls off much more rapidly than this, in particular at large
distances. This is mostly attributable to the SDSS spectroscopic
target selection. The chances of a candidate BHB star to be targeted
for spectroscopy in the course of SDSS and SEGUE is not a simple
function of its apparent magnitude, but depends on many (often
operational) factors, which have also evolved over the course of the
survey. The net result is that only under very favorable circumstances
are distant (hence faint) BHBs targeted, and have spectra obtained
that are of sufficient S/N to pass our quality criteria. As a result,
the radial distribution of our sample falls off much more steeply than
the parent population of halo BHB stars. We account for these effects
in the subsequent analysis, as described below.

Overall, our present sample of distant halo stars with available
kinematics is nearly an order of magnitude larger than that of
Battaglia et al. (2005), which has 240 stars. Because our tracers are
BHB stars, their distances are also known more accurately. However,
the Battaglia et al. sample does extend to larger radii, up to $\sim
100$~kpc.

%%%%%%%%%%%%%%%%%%%%%%%%%%%%%%%%%%%%%%%%%%%%%%%%%%%%%%%%%%%%%%%%%%%%%%%%%%%%%%%%%%%

\section{Modeling the BHB Kinematics}

We now describe our approach to convert the $\rm V_{los}(r)$
measurements (as a function of Galactocentric radius, r) for the $\sim
2400$ BHB stars into estimates of $\rm V_{cir}(r)$ of the Milky Way
halo, and ultimately to estimate $\rm M_{vir}$ for the Milky Way's
halo. With the full line-of-sight velocity data set, sampling radii of
$5-60$ kpc, we can obtain both the velocity dispersion, $\sigma_{\rm
  los}$, and identify a set of exceptionally high-velocity stars at
various radii that might be suitable to estimate $\rm V_{esc}(r)$.
Some previous work (e.g., Battaglia et al 2005) considered only
velocity dispersions, while others focused only on high-velocity stars
in the Solar neighborhood (Sakamoto et al. 2003; Smith et al. 2007).
We consider the full velocity distribution; as we shall see later, it
is close to a Gaussian and hence most of the pertinent information is
contained in $\sigma_{\rm los}$.

To link the observables, $\rm V_{los}$ and $\rm r$, to $\rm
V_{cir}(r)$, we must not only account for the particular survey
volume, but also need to make at least an implicit assumption about
the nature of the halo-star distribution function, in particular its
(an-)isotropy. In contrast to, e.g., Battaglia et al.  (2005), we
restrict ourselves not only to Jeans Equation modeling, but choose to
account for these issues also by investigation of a comparison with
cosmologically motivated galaxy simulations, to make ``mock
observations'' within these simulations, and then match test results
to the observations. One cannot expect the halo stars in the
simulations to have exactly the sample density profile as the actual
stars in the halo of the Milky Way; we account for such differences as
described below.

\subsection{Halo Star Kinematics in Simulated Galaxies}

Based on prior estimates of the halo mass, we chose two SPH
simulations corresponding to the formation of Milky Way-like haloes,
from which ``pseudo-observations'' are constructed for comparison with
the data. Both halos were picked from low-resolution cosmological dark
matter simulations and were re-simulated at higher resolution,
including the effects of gas dynamics, and star formation and stellar
feedback.

Simulation I was run using the smoothed particle hydrodynamics code
GRAPESPH (Steinmetz 1996), and included a moderately efficient stellar
feedback model in which the supernova energy was added to the thermal
energy of the gas, as described in Abadi et al. (2003). This simulated
halo has a present day virial mass of about $\sim 8.6 \times 10^{11}
~\rm M_\odot$ and a virial radius of $206$ kpc.

Simulation II was run using the smoothed particle hydrodynamics code
GADGET-2 (Springel 2005), assuming the standard star-formation
prescription of Springel et al. (2003), but without stellar feedback.
This simulated halo has a present-day virial radius of $\rm r_{vir} =
345$ kpc, and a virial mass of $\rm M_{vir}=2.1\times 10^{12}
M_{\odot}$ (these values differ from Naab et al. 2007, as we use a
different virial contrast in this paper).

For further details of the simulations we refer the interested reader
to Abadi et al. (2003, Simulation I) and to Naab et al. (2007,
Simulation II), and references therein.

\begin{figure}
\includegraphics[width=\textwidth]{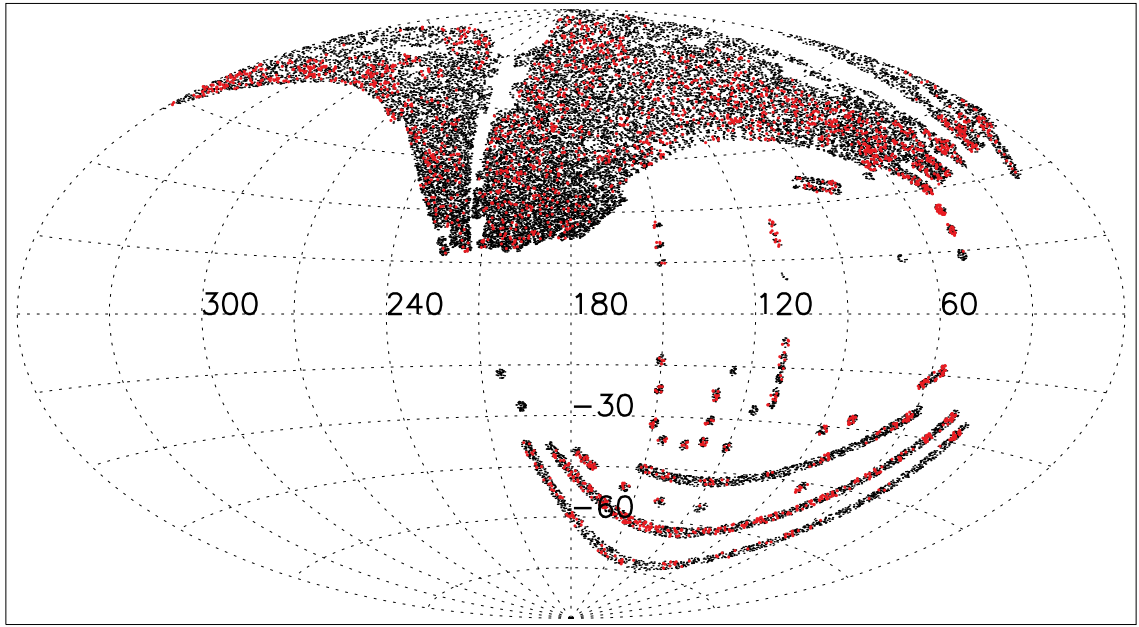}
\caption{The Galactic sky coverage of the observed BHB stars (red
  dots) and selected simulated stars (black dots), drawn from
  Simulation I.}
\label{f:f11}
\end{figure}

For each particle in these simulations we have the $3$-D positions and
$3$-D velocities, the circular velocity, $\rm V_{cir}\equiv \sqrt{ r
  \partial\Phi / \partial r}$, and the escape velocity, $\rm V_{esc}$,
to the virial radius of the simulation at each given position.
Because SDSS/SEGUE only observed $\approx$ $\pi$ steradians, not the
entire sky (see Figure~\ref{f:f11}), in order to compare the
observations to the simulations we must select simulated star
particles in the same region as the SDSS footprint.

As a first step, we specified ``Galactic coordinates'' in the
simulation by defining the ``Galactic Plane'' by the net angular
momentum of all (stellar) particles within 10~kpc of the center. In
this coordinate system one has (by definition) $\vec{L}_{tot} =
|{\mathrm L}_{tot}|\vec{e}_z$, and the Sun is at $\rm (x_\sun, y_\sun,
0.0)$ kpc ($\rm R_\sun=\sqrt{x_\sun^2+y_\sun^2}=8.0$ kpc). We can then
calculate the Galactocentric radial velocity as seen from the ``Sun''
for each simulation particle as:

\begin{eqnarray}  d=\rm \sqrt{(x-x_\sun)^2+(y-y_\sun)^2+z^2}\nonumber \\
  {\rm v}_{los}=\frac{\rm {v_x(x-x_\sun)+v_y(y-y_\sun)+v_zz}}{d}
\end{eqnarray}
Since the simulated ``observer'' is at rest in the Galactocentric
coordinate system, no assumptions about the LSR need be made in the
simulations.

To create ``pseudo-observations'', we first remove the satellites in
the simulation and then assume 10 positions for the Sun that all have
$\rm R~=~8~ kpc$ and $\rm z~=~0~ kpc$, but different azimuthal angles,
$\phi$. The corresponding $l$ and $b$ in this coordinate system can
then be used to select the star particles in the same region of the
simulation as our observed BHB sample. The effective flux limit of the
BHB sample, $g\leq 20$, implies a maximum distance from the Sun of
about $76$~kpc, a distance which we impose as a selection limit in the
simulations. To select ``halo'' stars, we also impose that the stars
have to be at least $4$~kpc above or below the disk plane. Finally, we
average these 10 samples of simulated halo stars. Figure~\ref{f:f11}
shows the distribution of these selected simulated stars and the
observed halo BHB stars in $l$ and $b$.

This procedure results in a sample of simulated halo stars (each
position produces an ``observational data set'') with Galactocentric
radial velocities, Galactocentric radii, escape velocities, and
circular velocities, whose distribution is shown in
Figure~\ref{f:f12}. This figure makes it clear that, even in a large
sample, the Galactocentric radial velocity rarely approaches the
escape velocity (one obvious reason being that we measure only the
projected component of the space velocity).

\subsection{Estimating $\rm V_{cir}(r)$ from the data}

We analyze the implications of the observed BHB kinematics for the
Milky Way halo's mass distribution in two steps.  We first estimate
$\rm V_{cir}(r)$ from the data/simulation comparison in a set of
statistically independent radial bins, effectively constructing a
circular velocity curve extending to 60 kpc.  We then fit the circular
velocity curve with NFW halo (and bulge+disk) models, resulting in
estimates of M$_{\rm vir}$.  It should be noted that the Milky Way
halo's presumed virial radius extends about a factor of four beyond
the most distant BHB stars in our observed sample of tracers.

\begin{figure}
\includegraphics[width=\textwidth]{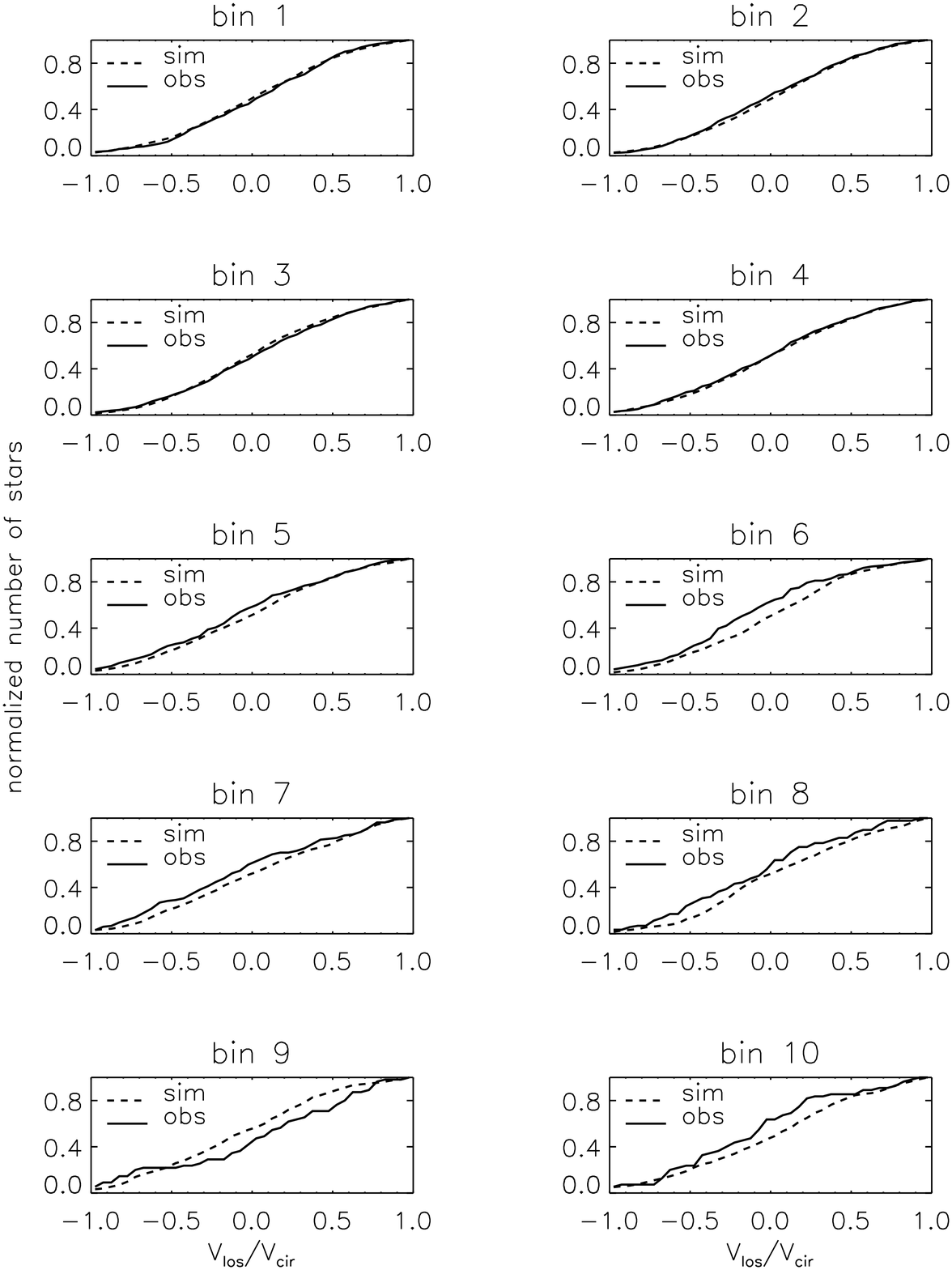}
\caption{Comparison of the Galactocentric line-of-sight velocity
  distribution, $\rm P(V_{los}/V_{cir})$ between the halo star
  particles in the Simulation I and the observations, shown here for
  all bins. The dashed line is $\rm P(V_{los,sim}/V_{cir})$, and the
  solid line is $\rm P(V_{los,obs}/V_{cir})$, after finding the best
  matching velocity scaling, $\rm V_{cir}$, listed in Table 3; the
  best match in this context is obtained from a K-S test.}
\label{f:f13}
\end{figure}

\begin{figure}
\includegraphics[width=\textwidth]{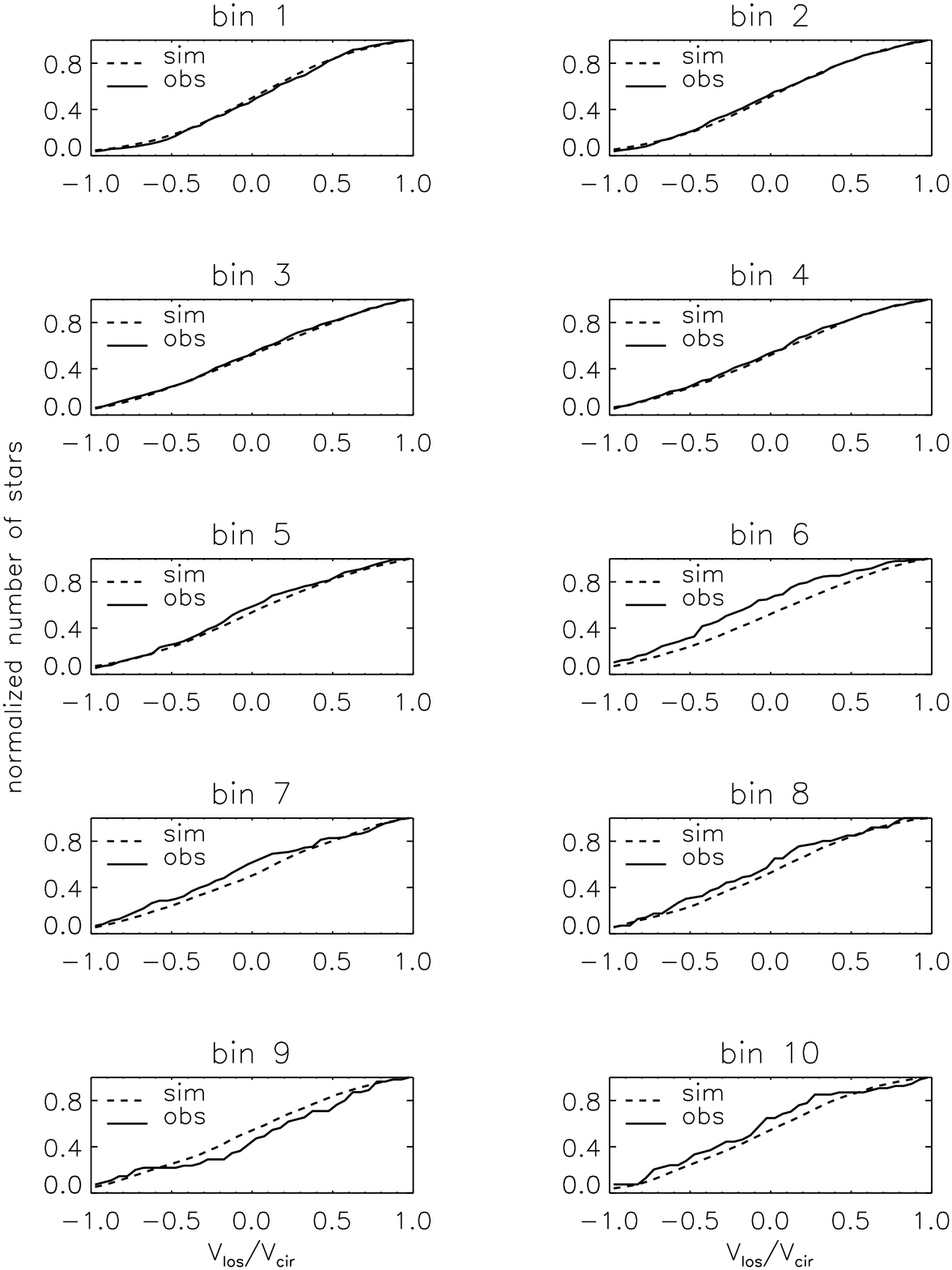}
\caption{Comparison of the Galactocentric line-of-sight velocity
  distribution, $\rm P(V_{los}/V_{cir})$ between the halo star
  particles in the Simulation II and the observations, shown here for
  all bins. The dashed line is $\rm P(V_{los,sim}/V_{cir})$, and the
  solid line is $\rm P(V_{los,obs}/V_{cir})$, after finding the best
  matching velocity scaling, $\rm V_{cir}$, listed in Table 3; the
  best match in this context is obtained from a K-S test.}
\label{f:f14}
\end{figure}

For the data/model comparison we construct the distributions, $\rm
P(V_{los}/V_{cir}(r)) $, for the simulations and the data
respectively. For the simulations, we use the procedure described in
the above section to obtain the $\rm V_{los}$ and $\rm V_{cir}$ at
each particle position. We compare those $\rm
P_{sim}(V_{los}/V_{cir}(r))$ distributions to analogous ones
constructed from the data for a sequence of trial values $\rm
V_{cir}(r)$. As the best observational estimate of $\rm V_{cir}(r)$,
we then take that value for which the probability that $\rm
P_{obs}(V_{los}/V_{cir})$ and $\rm P_{sim}(V_{los}/V_{cir})$ were
drawn from the same distribution is maximal. As we have no {\it
  a-priori} functional form for these distributions, we define this
best match as that which maximizes the probability in a two-sided K-S
test. To define confidence limits on the $\rm V_{cir}(r)$ estimate, we
repeat this procedure with bootstrapped versions of $\rm
P_{obs}(V_{los}/V_{cir})$, and take as $\rm \delta_{V_{cir}}$ the
variance of the resulting $\rm V_{cir}$ distribution. We divide the
Galactocentric radius ($\rm r$) into 10 bins and apply this method,
which results in 10 (statistically) independent radial estimates of
$\rm V_{cir}(r)$. Specifically, the bins we adopted were $5 - 10
$~kpc, $10 - 15 $~kpc, $15-20 $~kpc, $20-25 $~kpc, $25-30 $~kpc,
$30-35 $~kpc, $35-40 $~kpc, $40-45 $~kpc, $45-50 $~kpc, and $50-60
$~kpc (all distances are from the Galactic Center, not from the Sun).
The best matched distributions of $\rm P(V_{los}/V_{cir})$ are
illustrated in Figure~\ref{f:f13} and Figure~\ref{f:f14}. In these
figures, the red line shows $\rm P(V_{los}/V_{cir})$ of the
simulations, while the black line shows that of the data.  Overall,
the velocity distributions agree well with one another.

To estimate the statistical uncertainties on these values, we use
bootstrap resampling on our BHB sample (typically 100 times), repeat
the above procedure, and indicate the resulting 68\% (1-$\sigma$)
confidence region as error bars.

In a given gravitational potential, more centrally concentrated
kinematic tracer populations will exhibit smaller velocity dispersions
than more extended ones.  In the simple case of a spherical potential,
with tracers of $\rho \sim {\rm r}^{-\gamma}$ and an isotropic
velocity dispersion that varies only slowly with radius, the Jeans
Equation, for a given $\rm V_{cir}(r)$, yields the relative velocity
dispersions of the two populations, $\sigma_1 / \sigma_2 = \sqrt{
  \gamma_2 / \gamma_1}$, where the indices refer to two different
hypothetical tracer populations.

In the radial range of 10-60~kpc, the density profile of halo stars in
Simulation II is approximately $\rho \sim \rm r^{-2.9}$, while that of
Simulation I is $\rho \sim \rm r^{-3.7}$ . This, however, should not
be compared to the radial distribution of the stars for which we
actually have velocities, but to the radial profile of halo stars from
which they were drawn (see \S 2.3). This requires the reasonable
assumption that the measured velocities are uncorrelated with the
spectral targeting -- even a more complete sampling of BHB stars at a
given radius would have yielded the same velocity distribution. As an
estimate for the actual density profile for the Milky Way's stellar
halo at 10-60 kpc, we take the estimates (for main sequence turn-off
stars) of $\rho \sim \rm r^{-3.5}$ (Bell et al. 2007), based on SDSS.
Using the above correction based on the Jeans Equation, we must
subsequently revise the derived velocity scales for the Milky Way halo
upward by $\sqrt{3.5/2.9}=1.1$ for Simulation II and downward by
$\sqrt{3.5/3.7}=0.97$ for Simulation I.

\begin{figure}
\includegraphics[width=\textwidth]{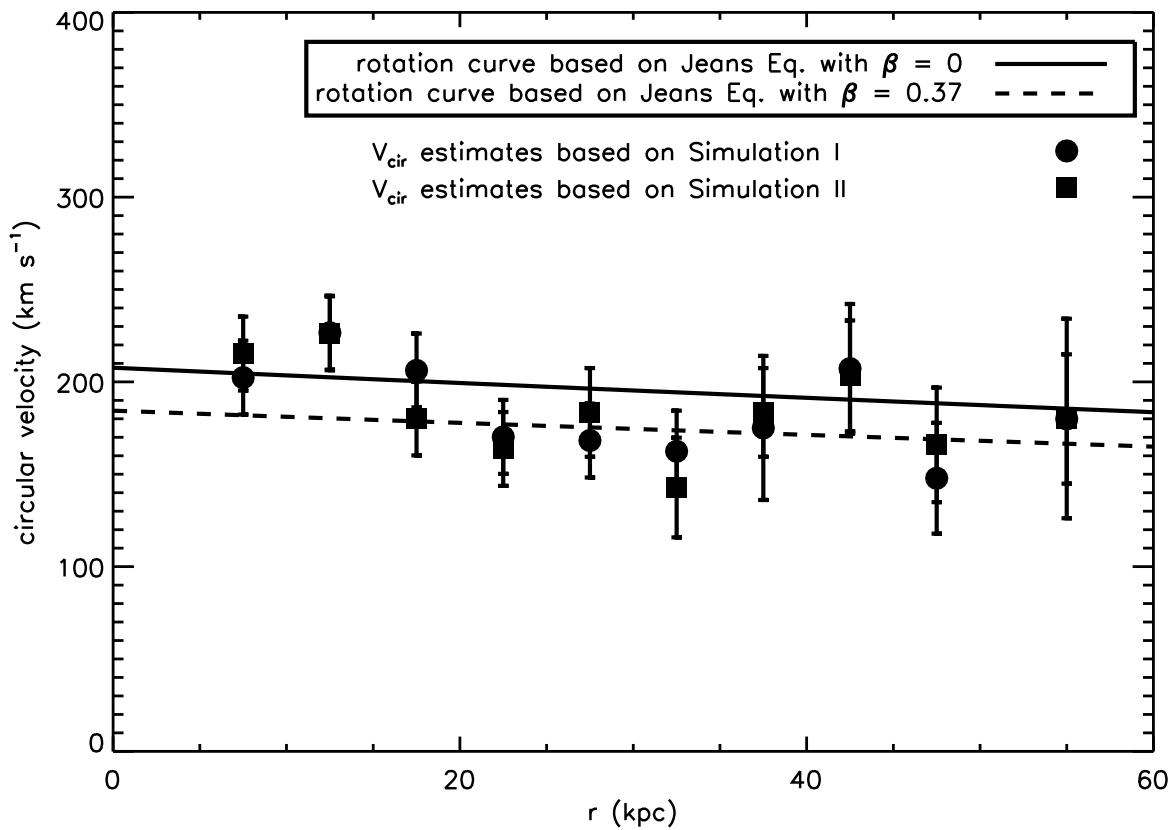}

\caption{The distribution of circular velocity estimates, V$\rm
  _{cir}$, for two different simulated galaxies. The filled circles
  are the V$\rm _{cir}$ estimates for the observed halo BHB stars
  based on Simulation I and the filled squares are the V$\rm _{cir}$
  estimates based on Simulation II. The two lines are the circular
  velocity curve estimates derived from the velocity dispersion
  profile (Figure~\ref{f:f10}) and the Jeans Equation with
  $\beta~=~0.37$ and $\beta~=~0$.}
\label{f:f15}
\end{figure}

Matching $\rm P(V_{los}/V_{cir}(r))$, and applying the above
correction, we obtain the estimates summarized in Figure~\ref{f:f15}
and in Table 3.  The filled circles in this Figure reflect the $\rm
V_{cir}(r)$ estimates based on Simulation I, while the filled squares
stand for the $\rm V_{cir}(r)$ estimates based on Simulation II. As
mentioned before, the error bars are from bootstrapping.

We have explored whether the derived estimates of $\rm V_{cir}(r)$
depend on our adopted value for the distance of the Sun from the
Galactic center, $\rm R_\odot$, or on the adopted local rotation
velocity, $\rm V_{lsr}$. If we vary $\rm R_\odot$ adopted in the
observation and simulations (see Eqn 4 and Eqn 7), from $7.5 ~\rm kpc$
to $8.5 ~\rm kpc$, $\rm V_{cir}(r)$ changes by only about 1\%. Our
results for the derived $\rm V_{cir}(r)$ are also insensitive to the
choice of the $\rm V_{lsr}$ adopted in the observation (see Eqn 5).
Taking $\rm V_{lsr}$ as $\rm 200 ~km~s^{-1}$ or $\rm 240 ~km~s^{-1}$
changes the estimated $\rm V_{cir}(r)$ by only 3\%.

For reference, we show how these estimates of $\rm V_{cir}(r)$ compare
to those derived from the Jeans Equation and the fit to $\sigma_{\rm
  los}(\rm r)$ shown in Figure~\ref{f:f10}. From the Jeans Equation,
$\rm V_{cir}(r)$ can be estimated from the velocity dispersion,
$\sigma_{\rm r}$ (Binney \& Tremaine 1987) as follows,

\begin{equation}
\rm - \frac{\rm r}{\rho}~\frac{\rm d(\sigma_{r}^2\rho)}{{\rm d}r}~-~2\beta\sigma_{r}^2=\rm V_{cir}^2(r)
\end{equation}
with\
\begin{equation}
\rm \beta=1-\frac{\sigma_{\rm t}^2}{\sigma_{\rm r}^2}
\end{equation}

where $\sigma_{\rm r}(\rm r)$ and $\sigma_{\rm t}(\rm r)$ are the
radial and tangential velocity dispersions, respectively, in spherical
coordinates, and $\rm \rho(\rm r)$ is the stellar density.

The distribution of the halo stars in the simulations are anisotropic,
with $\beta~=~0.37$, and the simulations exhibit $\rm
\sigma_{los}(r)~\approx~\sigma_{\rm r}(\rm r)$ for this particular
survey volume.  Taking from Figure~\ref{f:f10} the fit to $\rm
\sigma_{los}(r)$ of the BHB stars (Eqn 6), and assuming $\sigma_{\rm
  los}(\rm r)~\approx~\sigma_{\rm r}(\rm r)$ for BHB stars, we can
derive two circular velocity curves, for $\beta~=~0.37$ (anisotropic),
and $\beta~=~0$ (isotropic), respectively, by adopting $\rho(\rm
r)~\sim~\rm r^{-3.5}$ (see Figure~\ref{f:f15}). They both agree well
with the simulation-based estimates, and are not used any more in the
following analysis.
 %%%%%%%%%%%%%%%%%%%%%%%%%%%%%%%%%%%%%%%%%%%%%%%%%%%%%%%%%%%%
%%%%%%%%%%%%%%%%%%%

\section{Results}

\begin{figure}
\includegraphics[width=\textwidth,height=10cm]{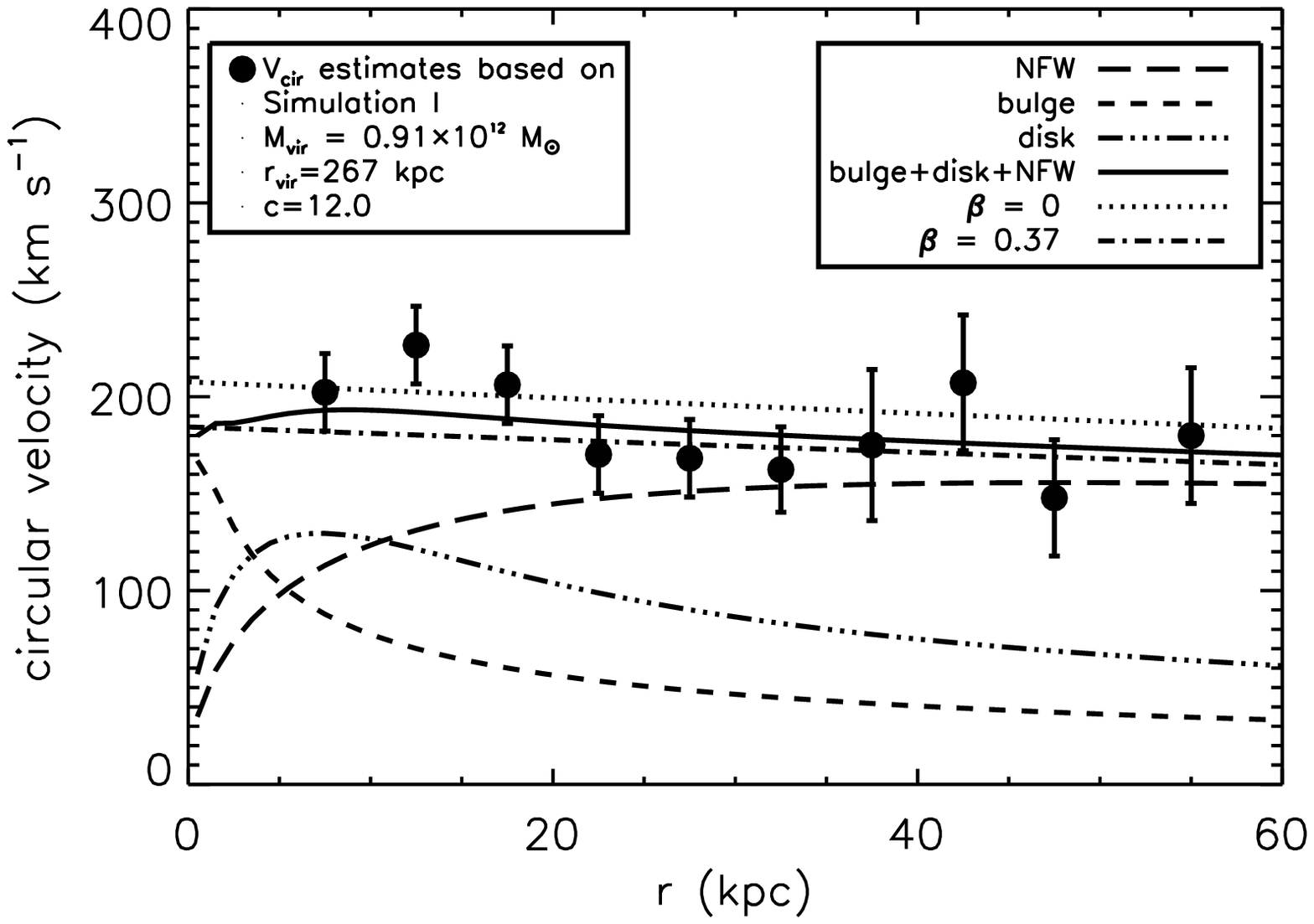}
\includegraphics[width=\textwidth,height=10cm]{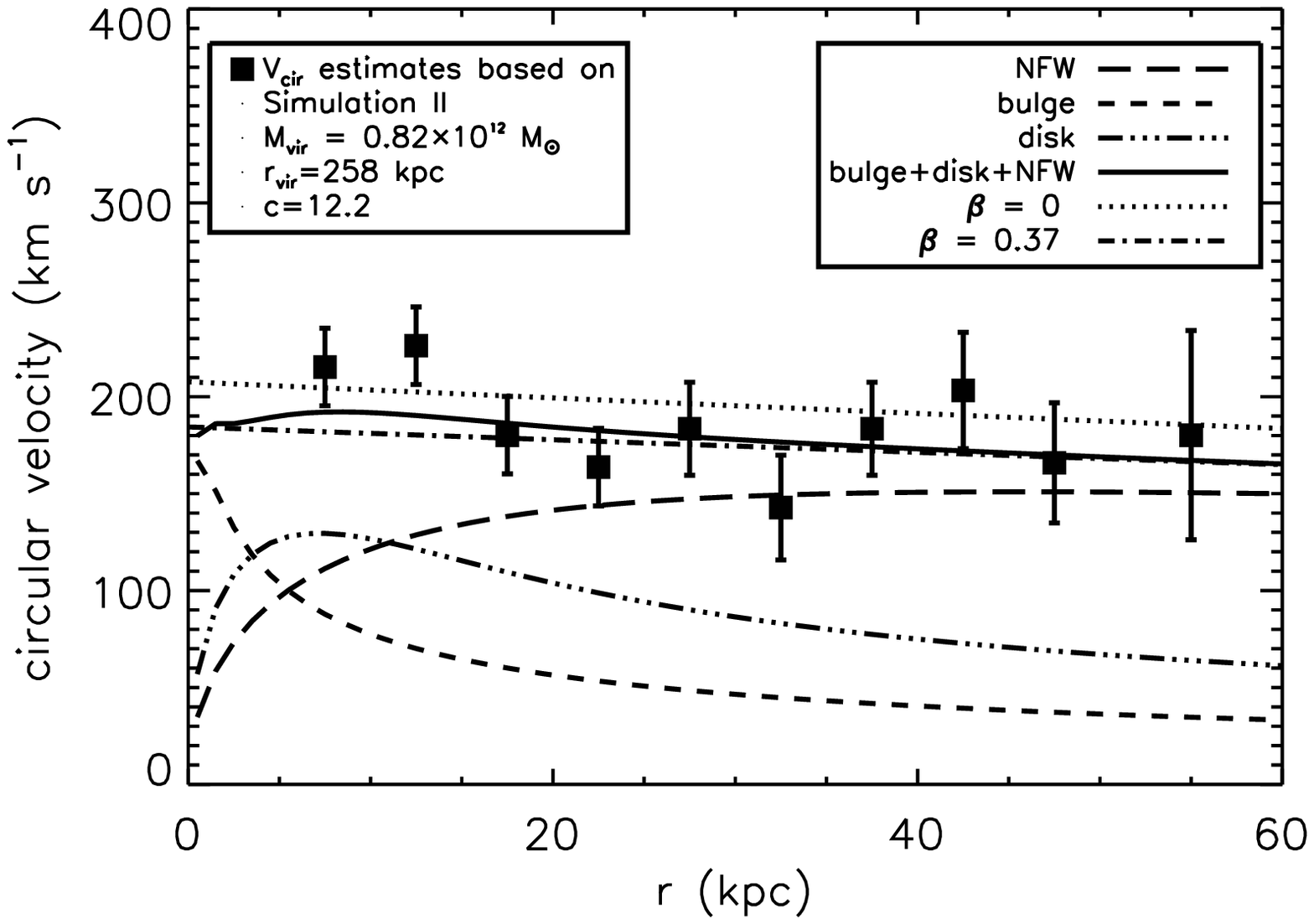}

\caption{Circular curve estimates matched by a combination of a
  stellar bulge, disk and an an unaltered NFW dark matter profile. The
  solid line is the best-fit circular velocity curve to the V$\rm
  _{cir}(r)$ estimates, while the large symbols in the two plots are
  the V$\rm _{cir}(r)$ estimates. Contributions of the adopted model
  components (i.e. disk, bulge, and halo) and the circular velocity
  curves based on the Jeans Equation are plotted in different
  linestyles. Estimates of virial mass, M$\rm _{vir}$, virial radius,
  r$\rm _{vir}$ and concentration parameter, $\rm c$, are labeled on
  the plots.}
\label{f:f16}
\end{figure}

\begin{figure}
\includegraphics[width=\textwidth,height=10cm]{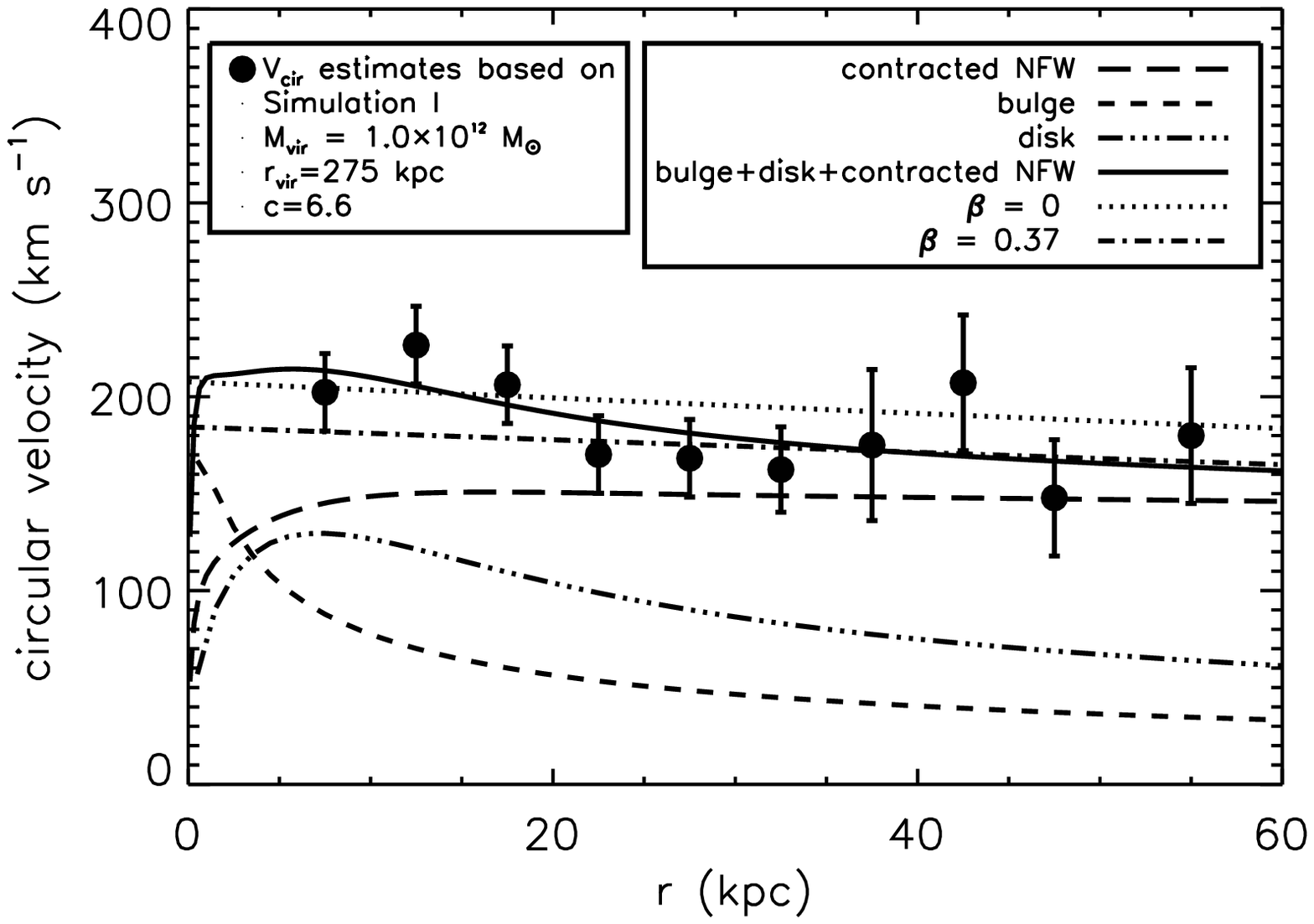}
\includegraphics[width=\textwidth,height=10cm]{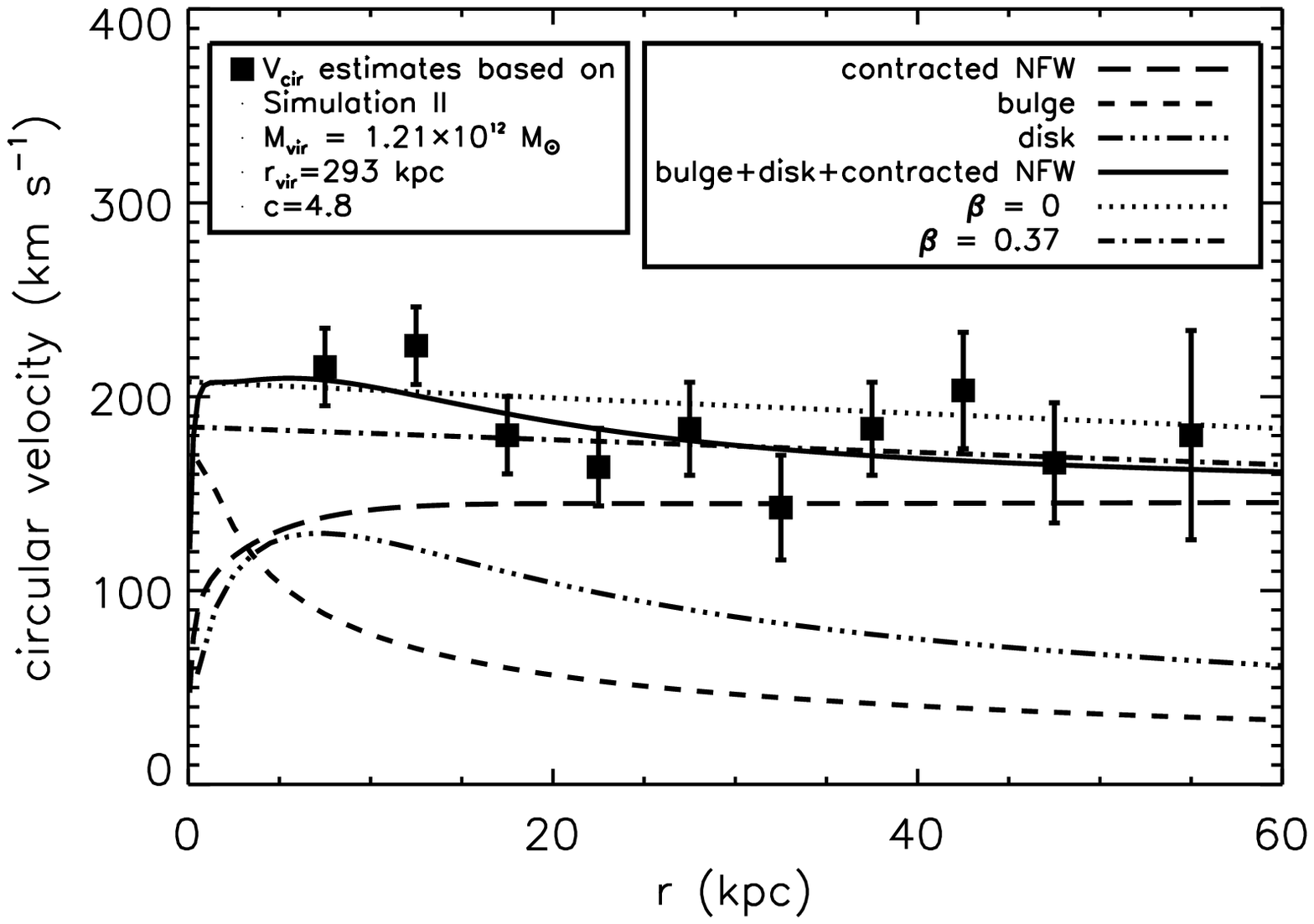}

\caption{As in Figure~\ref{f:f16}, but here the circular velocity
  curves were derived under the assumption of a contracted NFW
  profile. The solid line is the best-fit circular velocity curve to
  the V$\rm _{cir}(r)$ estimates, while the large symbols in the two
  plots are the V$\rm _{cir}(r)$ estimates. Contributions of the
  adopted model components (i.e. disk, bulge, and halo) and the
  circular velocity curves based on the Jeans Equation are plotted in
  different linestyles. Estimates of virial mass, M$\rm _{vir}$,
  virial radius, r$\rm _{vir}$ and concentration parameter, $\rm c$
  are labeled on the plots.}
\label{f:f17}
\end{figure}

Figure~\ref{f:f16} and Figure~\ref{f:f17} present the main result of
our analysis, $\rm V_{cir}(r)$, an estimate of the circular velocity
curve from $\sim 10-60$~kpc. This represents the first time that the
circular curve for the Galaxy has been estimated to such large
distances at this accuracy. Note that at small radii this estimate,
though derived from halo stars, agrees well (within $\sim 10$\%) with
established determinations at the Solar radius ($\sim 220 $ km $\rm
s^{-1}$). Beyond the Solar radius, the circular velocity curve appears
to be gently falling to 175 $\rm km~s^{-1}$ at $\sim 60$~kpc. Note
also that the circular velocity curve is a conceptually more robust
estimate than $\rm V_{esc}$, which depends more sensitively than V$\rm
_{cir}$ on $\rm \Phi (r)$ at radii beyond the measurements.

Using the functional form for $\rm V_{cir}(r)$ expected for an NFW
halo and the stellar component (see below) as a means to interpolate
the individual circular velocity curve estimates, one obtains $\rm
V_{cir}(60~{\rm kpc}) = 170\pm 15$ km $\rm s^{-1}$, or M$(<60~{\rm
  kpc}) = 4.0\pm 0.7 \times 10^{11}$M$_\odot$.  This is the largest
radius for which the data directly constrain $\rm V_{cir}(r) $ or $\rm
M(<r)$. Yet this radius is only one-fourth of the expected virial
radius of the Milky Way's halo. We therefore proceed with a separate
step, to use these $\rm V_{cir}(r)$ estimates to constrain
parameterized models for the overall dark matter halo. We assume that
the Galactic potential is represented by three components, a spherical
Hernquist (1990) bulge, an exponential disk for the stellar
components, and describe the halo by an NFW profile (Navarro et al.
1996). The total potential can then be simply expressed as

\begin{equation}
\Phi_{\rm tot}(\rm r)=\Phi_{\rm disk}(\rm r)+\Phi_{\rm bulge}(\rm r)+\Phi_{\rm NFW}(\rm r),
\end{equation} with an assumed potential, presumed to be spherically symmetric, for the disk and bulge of
\begin{equation}
\Phi_{\rm disk}(\rm r)=-\frac{\rm GM_{\rm disk}(1-e^{-\frac{\rm r}{\rm b}})}{\rm r},
\end{equation}
\begin{equation}
\Phi_{\rm bulge}(\rm r)=-\frac{\rm GM_{\rm bulge}}{\rm r+\rm c_0},
\end{equation} 
where $ \rm M_{\rm bulge}=1.5\times10^{10} \rm~ M_\sun$, $\rm c_0=0.6$
kpc, $\rm M_{disk}=5\times10^{10} \rm ~M_\sun$, and $\rm b=4$ kpc
(similar to Smith et al. 2007).  The radial potential for a spherical
NFW density profile can be expressed as
\begin{equation}
\Phi_{\rm NFW}(\rm r)=-\frac{\rm 4\pi G \rho_s r_{\rm vir}^3}{\rm c^3 r}{\rm ln} \left(1+\frac{\rm c r}{\rm r_{\rm vir}}\right),
\end{equation} 
where $\rm c$ is a concentration parameter, defined as the ratio of
the virial radius to the scale radius. For standard $\Lambda$CDM
cosmogonies we do not attempt to constrain halo flattening. The
parameter $\rho_{\rm s}$ is a characteristic density given by
\begin{equation}
\rho_{\rm s}=\frac{\rho_{\rm cr} \Omega_{\rm m} \delta_{\rm th}}{3}
\frac{\rm c^3}{\rm ln(1+c)-c/(1+c)},
\end{equation} 
where $\rho_{\rm cr}=3H^2/8\pi G$ is the critical density of the
universe, $\Omega_{\rm m}$ is the contribution of matter to the
critical density, and $\delta_{\rm th}$ is the critical overdensity at
virialization.  The virial mass can then be determined from the virial
radius, using
\begin{equation}
\rm M_{\rm vir}=\frac{\rm 4\pi}{\rm 3} \rho_{\rm cr} \Omega_{\rm m} \delta_{\rm
th} \rm r_{\rm vir}^3.
\end{equation} 
For our analysis we adopt $\Omega_{\rm m}=0.3 $, $\delta_{\rm
  th}=340$, and H$_o$ = 65$ ~\rm km ~s^{-1}$ ${\rm~ Mpc}^{-1}$.  Given
recent discussions (and doubts raised) of whether the baryons modify
the dark matter profile, as expected from ``adiabatic contraction''
(Dutton et al.  2007), we consider both an unaltered and an
adiabatically contracted NFW profile in the fit of $\Phi_{\rm tot}$.

By fitting the observed $\rm V_{cir}(r)$ with $\sqrt{ r {\rm d}\Phi /
  {\rm d}r }$ from $\rm \Phi_{tot}(r)$ shown as Eqn 10 we can
constrain the halo mass of the Milky Way. In this fit, we simply adopt
an unaltered NFW profile and a present-day relation between the mean
value of $\rm c$ and $\rm M_{vir}$,

\begin{equation}
\log_{10} {\rm c}=1.075-0.12(\log_{10} {\rm M_{vir}}-12).
\end{equation} 
This relation is accurate over the range $\rm 11~\leq ~\log {\rm
  M_{vir}} ~\leq~13$, and is based on the model of Macci\`{o} et al.
(2007) with $\Omega_{\rm m}=0.3$, $\Omega_\Lambda=0.7$,
$\sigma_8=0.9$, and $\rm n_i=1.0$.  Therefore, the $\rm M_{vir}$ is
derived as a 1-parameter fit (fit only $\rm M_{vir}$, presuming $\rm
c(M_{vir})$). The results are summarized in Figure~\ref{f:f16}.
  
Specifically, for the circular velocity estimates resulting from
Simulation I, we find
\begin{displaymath}
\rm M_{\rm vir}=0.91^{+0.27}_{-0.18}\times10^{12}\rm~M_\odot,
\end{displaymath} with $ \rm r_{\rm vir}=267^{+24}_{-19}$ kpc and $\rm c=12.0^{+0.3}_{-0.3}$.

For the circular velocity estimates based on Simulation II we find
\begin{displaymath}
 \rm M_{\rm vir}=0.82^{+0.21}_{-0.18}\times10^{12}\rm ~M_\odot,
\end{displaymath} with $ \rm r_{\rm vir}=258^{+20}_{-21}$ kpc and $\rm c=12.2^{+0.3}_{-0.4}$. 

Note that the error bars of $\rm M_{vir}$, $\rm r_{vir}$, and c are
determined from 1-$\sigma$ confidence intervals in a Chi-Square test.
For the two cases above we have adopted an unaltered NFW profile and
an average relation between $\rm c$ and $\rm M_{vir}$.  If we fit an
adiabatically contracted NFW profile (using the prescription of
Blumenthal et al. 1986 and Mo et al. 1998) and the same disk and bulge
as in Eqn 11 and Eqn 12, taking the concentration parameter ($\rm c$)
and virial mass (M$_{\rm vir}$) of the NFW profile as independent
parameters (i.e., we do not require that they follow the relationship
in Eqn 16), the $\rm M_{vir}$ can be derived as a 2-parameter fit
($\rm M_{vir}$ and $\rm c$), as shown in Figure~\ref{f:f17}.

For the circular velocity estimates resulting from Simulation I
\begin{displaymath}
\rm M_{\rm vir}=1.0^{+0.3}_{-0.2}\times10^{12}\rm M_\odot,
\end{displaymath} with $ \rm r_{\rm vir}=275^{+23}_{-20}$ kpc and $\rm c=6.6^{+1.8}_{-1.5}$.

The virial mass calculated by the circular velocity estimates based on
Simulation II is
\begin{displaymath}
\rm M_{\rm vir}=1.21^{+0.40}_{-0.30}\times10^{12}\rm M_\odot,
\end{displaymath} with $ \rm r_{\rm vir}=293^{+31}_{-26}$ kpc and $\rm c=4.8^{+1.2}_{-0.9}$.

The error bars on $\rm M_{vir}$ and $\rm r_{vir}$ are determined for
1-$\sigma$ confidence intervals in a Chi-Square test, fixing the
best-fit value of c, while the error bars on c are determined for
1-$\sigma$ confidence intervals in a Chi-Square test, fixing the
best-fit value of $\rm M_{vir}$.

Note that although contracted and uncontracted halo fits differ quite
strongly in their (initial) concentration, reassuringly the M$_{\rm
  vir}$ estimates remain relatively unaffected. The lower
concentrations for the contracted halo fits are still reasonally
consistent with the concentration scatter expected from cosmological
simulations (e.g., NFW 96).

The results here show that the Milky Way's circular velocity curve
must be gently falling to distances of 60 kpc from its value of $\sim
220$ km/s at the Solar radius; the null hypothesis $\rm V_{cir} (r)$
has a constant value of $220~km~s^{-1}$ is rejected by our fits with
very high statistical significance (at a level of 0.01).

A direct comparison with earlier work, at the data or $\rm
\sigma_{los}(r)$ level, is not straightforward to carry out, as each
sample has distinct selection effects, such as differing radial
distributions. Our estimate of $\rm M_{vir}$, taken to be the average
of the four estimates derived in this Section, $ 1.0^{+0.3}_{-0.2}
\times 10^{12}$M$_\odot$, falls well within the (considerably larger)
range of values estimated by Battaglia et al. (2005, 2006): $0.5 \sim
1.5 \times 10^{12}~{\rm M}_\odot$. This value is also reasonably
consistent with the recent estimate by Li \& White (2008), based on
Local Group dynamics. In general, however, the new mass estimate
presented here lies at the lower limit of most previous estimates.

The method to estimate $\rm V_{cir}(r)$ can also be used to derive the
escape velocity curve $\rm V_{esc} (r)$. We have carried out an
analogous procedure to that described in \S 3.2, using $\rm
P(V_{l.o.s}/V_{esc})$, and found lower limits on M$_{\rm
  vir}>0.5\times 10^{12} \rm M_{\sun}$, which do not appear to be as
stringent as the $\rm V_{cir}(r)$ estimates.

%%%%%%%%%%%%%%%%%%%%%%%%%%%%%%%%%%%%%%%%%%%%%%%%%%%%%%%%%%%%%%%%%%%%%%%%%%%%%%%%%

\section{Summary and Conclusion}

We have constrained the mass distribution of the Milky Way's dark
matter halo, by analyzing the kinematics of nearly $2400$ BHB stars
drawn from SDSS DR-6, which reach to Galactocentric distances of $\sim
\rm 60$ kpc. To obtain a ``clean sample'' of BHB stars, we have
re-analyzed all candidate BHB spectra, following the prescription by
Sirko et al. 2004, which should result in a contamination fraction
(mostly by BS stars) of well below 10\%. The metallicity distribution,
centered on [Fe/H] $ \sim -2$~dex, confirms that most sample members
must be halo stars. The distances to the BHB stars are known to $\sim
10\%$.

To account for the complex survey geometry and for plausible orbital
distributions of our sample of BHB stars, we have compared the
observed radial velocities to analogous quantities drawn from the
``star particles'' in galaxy formation simulations that resemble the
Milky Way. In particular, we have placed a virtual ``observer'' at 8.0
kpc from the simulation centers, looking in the actual SDSS directions
and sampling radial velocities for stars that are more than 4 kpc
above and below the disk plane. We then explored to what mass scale,
or $\rm V_{cir}(r)$, the simulations need to be scaled to in order to
match the observed line-of-sight velocity distribution in a set of
radial bins. In this analysis, we adjusted this scaling using the
Jeans Equation, to account for the slight difference in the radial
profile of observed halo stars ($\rm \rho \sim r^{-3.5}$) and
simulated halo stars ($\rm \rho \sim r^{-2.9}$) over this radial
range.

This procedure results in direct estimates of $\rm V_{cir}(r)$ from
$\sim 10$ kpc to $\sim 60$ kpc, the best such estimates to date over
this range. The circular velocity estimate varies slightly with
radius, dropping from $\sim 220~ \rm km~s^{-1}$ at 10~kpc to $\sim
170~ \rm km~s^{-1}$ in the most distant two bins. Applying this
procedure to two independent cosmological simulations (Simulation I
and Simulation II, respectively) results in consistent estimates of
$\rm V_{cir}(r)$. The mass enclosed within 60~kpc, constrained quite
directly by the data, is found to be $4.0\pm 0.7\times
10^{11}$M$_\odot$. As a result of the much more extensive data set
provided by SDSS, the uncertainties on this estimate are substantially
lower than those obtained by previous comparable work. A simple,
Jeans- Equation based modeling approach, assuming (an-)isotropies of
either $\rm \beta~ = ~0.37 ~or~ \beta~ = ~0$ (as found for the halo
stars in the cosmological simulations) yields results that are
consistent with these values.

Although each of the $\rm V_{cir}(r)$ points were estimated
independently, the implied overall profile is consistent with both the
mass profile in the simulations and with a parameterized mass model
that combines a fixed disk and bulge model with an NFW dark matter
halo, whose concentration c corresponds to the expected mean value for
its virial mass. We have also explored halo fits with the
concentration c as a free parameter, as well as halo profiles that
have been modified by adiabatic contraction. We find that our data
cannot discriminate well whether adiabatic contraction occurs or not
-- an uncontracted halo of higher than average concentration and a
contracted halo of (initially) low concentration fit comparably well.
The resulting virial masses, $\rm M_{vir}=1.0^{+0.3}_{-0.2} \times
10^{12}$M$_\odot$, are consistent for both fitting approaches. We have
also checked that these results are quite robust with respect to
distance errors, modest sample contamination ($\leq 10\% $), and the
choice of a different, independent galaxy formation simulation.

The estimate of $\rm M_{vir}$, which does imply an extrapolation from
r$_{\rm max}=60$ kpc to the virial radius of $\sim 250$ kpc, is
consistent with a recent estimate from a much smaller sample of halo
stars (Battaglia et al. 2006), but it is lower than previous estimates
that also rely on the kinematics of satellite galaxies (e.g., Kochanek
et al. 1996). However, recent results on the LMC (Kallivayalil et al.
2006; Besla et al. 2007) indicate that not even the Magellanic Clouds
may have been bound to the Milky Way for long, posing a potential
conceptual problems for the use of satellite dynamics. It should be
interesting to explore how our new constraint modifies the Local Group
timing argument and its implication for M31's halo mass (Li \& White
2008).

The estimate of $\rm M_{vir}\sim 10^{12}$M$_\odot$, together with an
estimated total cold baryonic mass of $6.5\times 10^{10}$M$_\odot$ and
a global baryon mass fraction of $0.17$, implies that nearly 40\% of
all baryons within the virial radius have cooled to form the Milky
Way's stars and (cold) gas, consistent with recent estimates for
galaxies of that mass scale, based on statistical arguments (van den
Bosh et al. 2007; Gnedin et al. 2007).

We note that parts of our analysis have been performed under the
assumption that the stellar halo of the Galaxy is considered as a
single relaxed population, or one that matches the simulations. Our
data on the overall dynamics are consistent with that hypothesis.
Recent evidence from Carollo et al. (2007) and Miceli et al. (2007)
suggest that the halo may well be more complex, and comprise two
distinct populations of inner- and outer-halo stars, with (slightly)
different net rotational velocities and spatial profiles. We defer all
analysis of kinematic and spatial sub-structure in this particular
sample to a future paper.

The SDSS and SEGUE surveys have shown in this context that they can
provide unparalleled sets of kinematic tracers for the Milky Way,
enabling a direct ``circular velocity curve'' estimate of the Milky
Way extending to 60 kpc. Once the full set of spectroscopy from SEGUE
is available, a much larger set of stars for such an analysis should
be available.  The proposed extension of SDSS, known as SDSS-III (and
using a more sensitive, 1000 fiber spectrograph), will provide the
opportunity to obtain higher quality spectra for fainter, more distant
BHB stars, thus extending the reach of our analysis to over 100 kpc.

\acknowledgments

We thank Dr.  Xianzhong Zheng and Dr.  Jianrong Shi for useful
discussions and assistance.

This work was made possible by the support of the Chinese Academy of
Sciences and the Max-Planck-Institute for Astronomy, and is supported
by the National Natural Science Foundation of China under grants Nos.
10433010 and 10521001.

TCB and YSL acknowledge partial support from the US National Science
Foundation under grants AST 06-0715 and AST 07-07776, as well as from
grant PHY 02-15783; Physics Frontier Center / Joint Institute for
Nuclear Astrophysics (JINA).

PRF acknowledges partial support through the Marie Curie Research
Training Network ELSA (European Leadership in Space Astrometry) under
contract MRTN-CT-2006-033481.

Funding for the Sloan Digital Sky Survey (SDSS) and SDSS-II has been
provided by the Alfred P.  Sloan Foundation, the Participating
Institutions, the National Science Foundation, the U.S. Department of
Energy, the National Aeronautics and Space Administration, the
Japanese Monbukagakusho, and the Max Planck Society, and the Higher
Education Funding Council for England.  The SDSS Web site is
http://www.sdss.org/.  The SDSS is managed by the Astrophysical
Research Consortium (ARC) for the Participating Institutions.  The
Participating Institutions are the American Museum of Natural History,
Astrophysical Institute Potsdam, University of Basel, University of
Cambridge, Case Western Reserve University, The University of Chicago,
Drexel University, Fermilab, the Institute for Advanced Study, the
Japan Participation Group, The Johns Hopkins University, the Joint
Institute for Nuclear Astrophysics, the Kavli Institute for Particle
Astrophysics and Cosmology, the Korean Scientist Group, the Chinese
Academy of Sciences (LAMOST), Los Alamos National Laboratory, the
Max-Planck-Institute for Astronomy (MPIA), the Max-Planck-Institute
for Astrophysics (MPA), New Mexico State University, Ohio State
University, University of Pittsburgh, University of Portsmouth,
Princeton University, the United States Naval Observatory, and the
University of Washington.

%%%%%%%%%%%%%%%%%%%%%%%%%%%%%%%%%%%%%%%%%%%%%%%%%%%%%%%%%%%%%%%%%%%%%%%%%%%%%%%

\clearpage

%%%%%%%%%%%%%%%%%%%%%%%%%%%%%%%%%%%%%%%%%%%%%%%%%%%%%%%%%%%%%%%%%%%%%%%%%%%%%%%%%%%%%%
%%Figure

%%%%%%%%%%%%%%%%%%%%%%%%%%%%%%%%%%%%%%%%%%%%%%%%%%%%%%%%%%%%%%%%%%%%%%%%%%%%%
%%Table
\begin{deluxetable}{cccccccccccccccccccc}
\tablecaption{List of $10224$ BHB candidates selected by color cut}
\label{t:tbl1}
\tablecolumns{20}
\small
\tabletypesize{\scriptsize}
\tablewidth{0pt}
\tablehead{\colhead{$\rm SpName$} & \colhead{$\rm Name$} & \colhead{$\rm RA $} & \colhead{$\rm  Dec$ }& \colhead{$\rm l$ }& \colhead{$\rm b$ }& \colhead{$\rm g$} & \colhead{$\rm g-r$} & \colhead{$\rm u-g$ }& \colhead{uerr} & \colhead{gerr} & \colhead{rerr} &\colhead{HRV} & \colhead{HRVerr} & \colhead{RVgal} & \colhead{$\rm f_{m,\delta}$} & \colhead{$\rm D_{0.2,\delta}$} & \colhead{$\rm c_\gamma$} & \colhead{$\rm b_\gamma$} & \colhead{$\rm Type$}\\
\colhead{$--$} & \colhead{$--$} & \colhead{degree} & \colhead{degree} & \colhead{degree} & \colhead{degree} & \colhead{mag} & \colhead{ mag} & \colhead{mag} & \colhead{mag} & \colhead{mag} & \colhead{mag} & \colhead{$\rm km~s^{-1}$} & \colhead{$\rm km~s^{-1}$} & \colhead{$\rm km~s^{-1}$} & \colhead{$---$} & \colhead{\AA} &\colhead{ $--$} & \colhead{\AA} &\colhead{ $--$}}

\startdata
51602-0266-225 & SDSSJ094218.23-002519.7 & 145.575943 & -0.422125 & 236.196579 & 36.896130 & 15.46 & -0.10 & 1.20 & 0.02 & 0.02 & 0.01 & -95.7 & 2.1 & -245.5 & 0.24 & 26.93 & 0.84 & 9.97 & BHB\\
51602-0266-397 & SDSSJ094138.17+001821.5 & 145.409058 & 0.305967 & 235.328827 & 37.175690 & 18.48 & -0.11 & 1.22 & 0.04 & 0.02 & 0.02 & -36.0 & 12.8  & -183.8 & 0.20 & 27.78 & 1.12 & 10.57 & BHB\\
51602-0266-634&SDSSJ094840.23+002818.0&147.167633&0.471673&236.434402& 38.711887&18.01&-0.02&1.02&0.03&0.02&0.01&9.4&10.5 &-136.8&0.29 &25.80 &0.92&10.01&BHB\\
51609-0292-102&SDSSJ125223.54-003708.2&193.098083& -0.618937& 303.444397&62.251862 &14.29 & -0.24 & 1.27 & 0.02& 0.02&0.02   & 227.3  & 1.6 & 148.7 & 0.23 & 27.00 &1.03  & 10.16 &BHB\\
51609-0292-155&SDSSJ125050.87-000806.1&192.711960 &-0.135032& 302.609894&62.736347&17.84 & -0.00& 1.06&0.02 &0.02&0.02 & 190.8& 7.2&112.8 &0.24 &27.46 & 0.85 &10.27 &   BHB\\
51609-0292-232& SDSSJ124759.81-000456.2&191.999207 &-0.082266 &301.051117 &62.776917&18.16& -0.04&1.05 &0.03&0.02 &0.02 & -121.4&  9.3 & -200.9 & 0.28 &  30.32 &  0.86 & 10.31 &BS\\
51609-0292-269 &SDSSJ124721.11-002931.5  &191.837952 & -0.492089 &300.729431 &62.362186 &19.17  &-0.09&1.12 &0.05 &0.03& 0.02 &232.0 & 19.5& 151.0 & 0.30 &32.62 &0.93& 10.10 & MS\\
51609-0292-329&SDSSJ124641.66+003751.2&191.673569& 0.630884 &300.275543&63.478165&17.15&-0.10&1.23& 0.02 &0.01&0.02& -49.0& 3.9 & -127.2 & 0.23 &25.03 & 0.81&8.25&BHB\\
51609-0292-351&SDSSJ124449.35+002157.4&191.205612 &0.365958& 299.263092&63.190571& 17.64 &-0.26& 1.13& 0.02 & 0.01& 0.01 &4.2 &5.7&-75.8 & 0.24& 24.54 &1.15  & 9.51 &   BHB\\
51609-0292-367& SDSSJ124805.12+010113.5 & 192.021347&1.020428 &301.028198  &63.879784&16.38 & -0.17 &  1.22 &  0.03& 0.02 & 0.02& 84.8 &2.8 & 8.6 & 0.24  &  34.34  &0.87 &11.53 &BS\\
\enddata
\tablecomments{The first two columns are object names and the next
  four columns contains the astrometry (ra, dec, l, b) for each
  object.The magnitude and color are in the next six columns:
  corrected for extinction. The radial velocities and errors are
  listed next. The next four columns are the linewidth parameters from
  the Balmer lines. The last column is the classification by $D_{0.2}$
  \& $f_m$ method. The complete version of this table is in the
  electronic edition of the Journal. The printed edition contains only
  a sample.}
\end{deluxetable}

\begin{deluxetable}{cccccccccccccccccccccccccc}
\tablecaption{List of $2558$ BHB stars selected from SDSS DR6}
\label{t:tbl2}
\tablecolumns{26}
\small
\tabletypesize{\scriptsize}
\tablewidth{0pt}
\tablehead{\colhead{$\rm SpName$} & \colhead{$\rm Name$} & \colhead{$\rm RA $} & \colhead{$\rm  Dec$ }& \colhead{$\rm l$ }& \colhead{$\rm b$ }& \colhead{$\rm g$} & \colhead{$\rm u-g$} & \colhead{$\rm g-r$ }& \colhead{$M_g$} &  \colhead{$\rm D_{0.2,\delta}$} & \colhead{$\rm f_{m,\delta}$} & \colhead{$\rm c_\gamma$} & \colhead{$\rm b_\gamma$} & \colhead{d} &\colhead{r} &\colhead{x} &\colhead{y} &\colhead{z} &\colhead{HRV} & \colhead{HRVerr} & \colhead{RVgal} &  \colhead{$T_{eff}$} &\colhead{$\log g$} & \colhead{[Fe/H]}&\colhead{$\rm Type$}\\
\colhead{$--$} & \colhead{$--$} & \colhead{degree} & \colhead{degree} & \colhead{degree} & \colhead{degree} & \colhead{mag} & \colhead{ mag} & \colhead{mag} & \colhead{mag} & \colhead{\AA} & \colhead{$--$} & \colhead{$--$} & \colhead{\AA} &\colhead{kpc} &\colhead{kpc}&\colhead{kpc}&\colhead{kpc}&\colhead{kpc}&\colhead{$\rm km~s^{-1}$} & \colhead{$\rm km~s^{-1}$} & \colhead{$\rm km~s^{-1}$}  & \colhead{K} &\colhead{$--$} & \colhead{$--$} &\colhead{ $--$}}

\startdata
51602-0266-125 &    SDSSJ094317.57-011021.2   &       145.823196 &          -1.172550 &         237.138931 &          36.662491&   16.48 &   1.17  & -0.22  &  0.55  & 26.60   & 0.21 &   1.02  & 10.07&    15.3 &   20.2 &   14.7 &   10.3&     9.2 &  178.0 &    3.5  &  26.2 &   8315  &  3.12 &  -2.03 &  BHB\\
51602-0266-397&     SDSSJ094138.17+001821.5&          145.409058&            0.305967&          235.328827 &          37.175690&   18.48&    1.22 &  -0.11&    0.55&   27.78&    0.20&    1.12&   10.57&    38.5&    42.8&    25.5&    25.3&    23.3&   -36.0&    12.8&  -183.8&    8306&    3.81&   -1.97 &  BHB\\
51602-0266-634 &    SDSSJ094840.23+002818.0 &         147.167633  &          0.471673&          236.434402&           38.711887  & 18.01&    1.02 &  -0.02 &   0.60 &  25.80  &  0.29  &  0.92&   10.01&    30.3 &   34.6 &   21.1 &   19.7&    19.0 &    9.4 &   10.5 & -136.8 &   7850 &   4.20&   -1.27&   BHB\\
51609-0292-102&     SDSSJ125223.54-003708.2  &        193.098083 &          -0.618937 &         303.444397 &          62.251862 &  14.29 &   1.27 &  -0.24&    0.55 &  27.00 &   0.23 &   1.03&   10.16&     5.6&     8.5 &    6.6&     2.2&     5.0 &  227.3 &    1.6 &  148.7 &   8668 &   2.61 &  -1.96 &  BHB\\
51609-0292-329 &    SDSSJ124641.66+003751.2  &        191.673569     &       0.630884  &        300.275543&           63.478165&   17.15 &   1.23  & -0.10&    0.55 &  25.03&    0.23 &   0.81 &   8.25 &   20.9&    20.6 &    3.3 &    8.1  &  18.7&   -49.0  &   3.9 & -127.2 &   7596 &   2.95&   -2.00 &  BHB\\
51609-0292-351  &   SDSSJ124449.35+002157.4 &         191.205612 &           0.365958  &        299.263092  &         63.190571 &  17.64 &   1.13  & -0.26 &   0.60 &  24.54&    0.24  &  1.15 &   9.51 &   25.6 &   25.1 &    2.4 &   10.1 &   22.8  &   4.2   &  5.7  & -75.8 &   9247 &   3.11&   -1.43&   BHB\\
51609-0292-582  &   SDSSJ125254.01+002903.2  &        193.225037  &          0.484220 &         303.747009 &          63.353657 &  16.22   & 1.23  & -0.13&    0.55 &  26.21  &  0.25&    0.85 &   9.09  &  13.6  &  14.0 &    4.6  &   5.1 &   12.2  & -42.9  &   3.2  &-118.0&    7953&    3.14 &  -1.66&   BHB\\
51609-0304-219  &   SDSSJ141723.88-002220.8&          214.349503 &          -0.372432 &         343.356750&           55.602268 &  16.64 &   1.05 &  -0.31 &   0.70   &23.05  &  0.20  &  1.19 &   9.68 &   15.4 &   13.0 &   -0.3 &    2.5 &   12.7 &  -25.0 &    3.8 &  -50.1  & -9999 &  -9.99&   -9.99&   BHB\\
51613-0305-243  &   SDSSJ142408.24-002930.7 &         216.034348   &        -0.491848 &         345.608093 &          54.511181 &  15.71&    1.17 &  -0.02 &   0.60 &  25.01  &  0.25 &   0.87 &   8.94 &   10.5 &    8.9 &    2.1  &   1.5&     8.6  & -78.4 &    2.4&   -99.4&    7754 &   3.68 &  -2.26&   BHB\\
51613-0305-488   &  SDSSJ142826.28+002915.4 &         217.109497&            0.487611&          348.109955 &          54.617874 &  17.54  &  1.17  & -0.26  &  0.60  & 24.71 &   0.22 &   1.12 &   9.55&    24.4&    21.0 &   -5.8&     2.9&    19.9 &   74.7&     4.2&    59.4 &   9338 &   2.87 &  -2.95 &  BHB\\
51614-0281-438&     SDSSJ112513.99+004207.5&          171.308289 &           0.702082 &         261.294983 &          56.438610 &  18.51 &   1.13 &  -0.23 &   0.60 &  24.60  &  0.22  &  1.16  &  9.44  &  38.2 &   39.7  &  11.2  &  20.9 &   31.8 &  252.4  &   9.6 &  134.5 &   8609  &  3.13  & -1.56&   BHB\\
\enddata
\tablecomments{The first two columns are object names and the next
  four columns contains the astrometry (ra, dec, l, b) for each
  object.The magnitude and color are in the next four columns:
  corrected for extinction. The next four columns are the linewidth
  parameters from the Balmer lines. The positions are listed in the
  next five columns. The radial velocities and errors are listed next.
  The next three columns are atomospheric parameters estimated by Ron
  Wilhelm.  The last column is the classification by combination of
  the $D_{0.2}\ vs.\ f_m$ method and the {\it scale width vs. shape}
  method. The complete version of this table is in the electronic
  edition of the Journal. The printed edition contains only a sample.}
\end{deluxetable}

\begin{deluxetable}{lcc}
\tablecaption{Estimates of the Circular Velocities as a Function of Distance}
\label{t:tbl3}
\tabletypesize{\scriptsize}
\tablewidth{0pt}
\tablehead{\colhead{r ~(kpc)}& \colhead{$\rm V_{cir,I}~$ (km s$^{-1})$} & \colhead{$\rm V_{cir,II}~ $(km s$^{-1})$}}
\startdata
7.5  & $202~\pm~20$  & $215~\pm~20$ \\
12.5 & $227~\pm~20$ & $226~\pm~20$ \\ 
17.5 & $206~\pm~20$  & $180~\pm~20$ \\
22.5 & $170~\pm~20$  & $164~\pm~20$ \\
27.5 & $168~\pm~24$ & $183~\pm~20$ \\
32.5 & $162~\pm~27$  & $143~\pm~22$ \\
37.5 & $175~\pm~24$  & $183~\pm~39$ \\
42.5 & $207~\pm~30$  & $203~\pm~35$ \\
47.5 & $148~\pm~31$  & $166~\pm~30$ \\
55.0 & $180~\pm~54$  & $180~\pm~35$ \\
\enddata
\tablecomments{ V$\rm_{cir,I}$ denote the estimates based on
  Simulation I, while V$\rm_{cir,II}$ denote the estimates based on
  Simulation II. These estimates have been derived from the velocity
  scaling of the observed distribution that yields the best agreement
  to the simulated velocity distribution, as determined by a K-S test
  probability, and have been corrected for the effects of different
  halo stellar densities in the observations and simulations. See \S
  3.2 for further information.}
\end{deluxetable}

\end{document}